\documentclass[%
 reprint,10pt,onecolumn,
 amsmath,amssymb,
 aps,superscriptaddress
]{revtex4-1}

\usepackage{textcomp}
\usepackage{amsmath}
\usepackage{amssymb}
\usepackage{graphicx}
\usepackage{color}
\usepackage{soul}
\usepackage{array}

\usepackage{titlesec}

\usepackage{amsfonts,graphics,dcolumn,bm,enumerate,amsthm}
\usepackage{comment,natbib,appendix,mathtools}
\usepackage{float}
\usepackage{hyperref} %Add a hyperlink in bibtex

\newcommand{\ac}
{\affiliation{Department of Computer Science, Asutosh College, Kolkata 700026, India}}
\newcommand{\sinp}
{\affiliation{Saha Institute of Nuclear Physics, Kolkata 700064, India}}
\newcommand{\snb}
{\affiliation{S. N. Bose National Centre for Basic Sciences, Kolkata 700106, India}}
\newcommand{\isi}
{\affiliation{Economic Research Unit, Indian Statistical Institute, Kolkata 700108, India}}

\begin{document}

%=================================================================
% Full title of the paper (Capitalized)
\title{Development of Econophysics: A Biased Account From Kolkata}

\author{Bikas K Chakrabarti}
 \email{bikask.chakrabarti@saha.ac.in}
 \sinp \snb \isi
 
 \author{Antika Sinha}
 \email{antikasinha@gmail.com}
 \ac \sinp

% Abstract 
\begin{abstract}
We present here a somewhat personalized account
of the emergence of econophysics as an
attractive research topic in physical as well
as social sciences. After a rather detailed
story telling about our endeavors from Kolkata,
we give a brief description of the main research
achievements in a simple and non-technical language. 
We also present briefly, in technical language,
a piece of our recent research result. We
conclude our paper with a brief perspective.

[The original version is already published (Feb.
2021) in Entropy vol. 23, art. 254 (Open online:
https://www.mdpi.com/1099-4300/23/2/254/pdf).
This version is slightly modified and updated
(Aug. 2021)].
\end{abstract}

\maketitle

\section{Introduction}

\begin{figure*}[htb]
\begin{center}
\includegraphics[width=8.5cm]{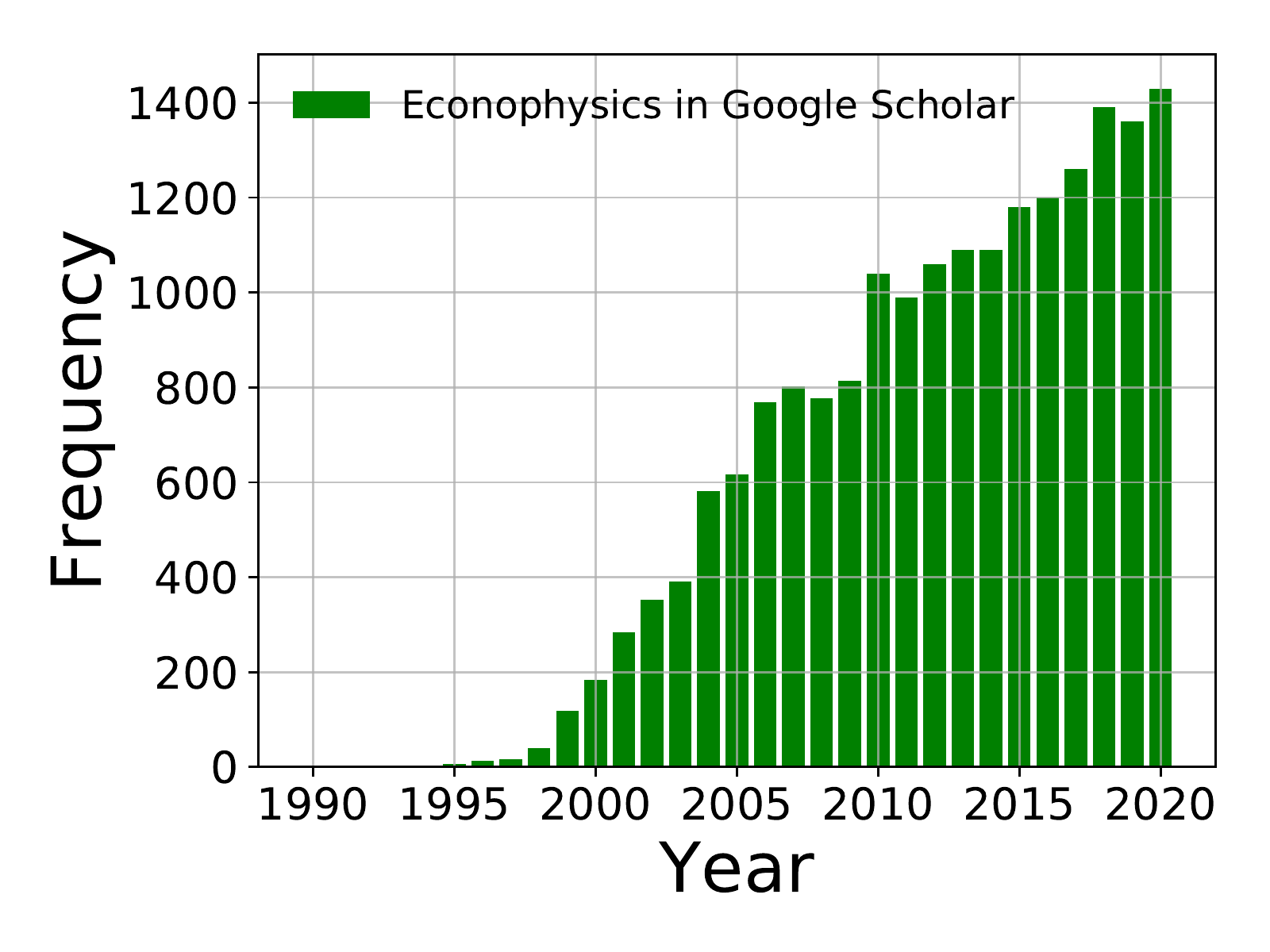} 
\end{center}
\caption{ Histogram plot of year-wise numbers of entries
containing the term econophysics against the
corresponding year. The data are taken from google
scholar (dated 28th August, 2021). It may also be
noted from Google Scholar that while this 25-year
old econophysics has today typical yearly citation
frequency of order $1.4 \times 10^3$, more than
100-year old subjects like astrophysics (Meghnad
Saha published his thermal ionization equation
for solar chromosphere in 1920), biophysics (Karl
Pearson coined the term in his 1892 book `Grammar
of Science') and geophysics (Issac Newton explained
planetary motion, origin of tides, etc in `Principia
Mathematica', 1687) today (28th August, 2021) have
typical yearly citation frequencies of order
$64.0 \times 10^3$, $34.3 \times 10^3$, and
$38.9 \times 10^3$ respectively.
 }
\label{fig_secI_I}
\end{figure*}

Countless attempts and researches, mostly in physics, to model and
comprehend the economic systems are about a century old. For the last
three or four decades, major endeavor have been made and some  successes
have been achieved and published, notably under the general title
`Econophysics'. The term was coined at a Kolkata conference held in 1995
by Eugene Stanley, who later in an  interview said `` ... So he (Bikas)
started to have meetings on econophysics and I think the first one was
probably in 1995 (he decided to start it in 1993–1994). Probably the first
meeting in my life on this field that I went to was this meeting. In that
sense Kolkata is — you can say — the nest from which the chicken was born
...''~\cite{gangopadhyay2013interview}. The entry on Econophysics by Berkeley Rosser in
the New Palgrave Dictionary of Economics (2nd Edition~\cite{rosser2008econophysics}) starts with the
sentence ``According to Bikas Chakrabarti (...), the term `econophysics'
was neologized in 1995 at the second Statphys-Kolkata conference in
Kolkata (formerly Calcutta), India, by the physicists H. Eugene Stanley
... See also Fig.~\ref{fig_secI_I} (and~\cite{stanley2000introduction}). It may be mentioned here that in
a more generalized sense, the term `Sociophysics' was introduced more than a
decade earlier by Serge Galam and coworkers~\cite{galam1982sociophysics} (see also~\cite{galam2012sociophysics}).''

 As we will discuss in the next section, economics, like all the natural
sciences (physics, chemistry, biology, geology, etc) are,
epistemologically speaking, knowledge or truth acquired through
induction  from observations (natural or controlled in the
laboratories) using inductive logic and analyzed or comprehended using
deductive logic (like mathematics). The divisions of natural sciences
between the streams like physics, chemistry, biology, geology are for
convenience and are man-made. `Truth' established in one branch or stream
of natural science does not become `false' or wrong in another; only the importance
often vary. This helps in the growth of an younger branch of science
through interdisciplinary fusion of established knowledge from another
older established branch; astrophysics, geophysics, biophysics and
biochemistry had been earlier examples. Econophysics has been the latest
one, and this special issue of Entropy attempts to capture the history,
success and future prospect of econophysics researches.

In the next five sections (essentially following
the structure and section-titles, suggested by the
Editors of this special issue) we discuss in
non-technical language, allowing them to be
accessible to the uninitiated younger students and
researchers (except in section 4, where we present
some new results of our research). True to
the spirit suggested by the Editors, the second
section has been presented in the form of a
`Dialogue' using the format of questions and answers
between us, the coauthors.

\section{What Attracted You to Econophysics?}

As mentioned earlier, this section is formatted in the form of a
dialogue (question and answer) between the coauthors.

\noindent
AS: What attracted you to Econophysics?  Can you briefly
recount? 

\noindent
BKC:  Meghnad Saha, founder of Saha Institute
of Nuclear Physics (named so after his death in 1956), had
been a pioneering Astrophysicist (known for the Saha
Ionization formula in astrophysics), had also thought
deeply about the scientific foundation of many social
issues (see e.g.,~\cite{chakrabarti2018econophysics}). In early seventies, our
undergraduate-level text book on heat and thermodynamics
had been `Treatise on Heat'~\cite{saha1931treatise}, written by Saha
together with Biswambhar Nath Srivastava (first published
in 1931). This turns out to be the earliest textbook
where the students were encouraged, in the section on
Maxwell-Boltzmann distribution in kinetic theory of ideal
gas, to compare it with the anticipated `gamma' distribution
of income in the society (see the page from~\cite{saha1931treatise} reproduced
in~\cite{chakrabarti2018econophysics}). If taken seriously, it asks the students to model
the income distribution in a society, which maximizes the
entropy (assuming stochastic market transactions)!\\

\noindent
AS: Before you go further, let me ask why should one think of applying
statistical physics to society in the first place?

\noindent
BKC: One Robinson Crusoe in an island can not develop a running
market or a functional society. A typical thermodynamic system,
like a gas, contains Avogadro number (about $10^{23})$ of
atoms (or molecules). Compared to this, the number ($N$ ) of
`social atoms' or agents in any market or society is of course
very small (say, about $10^2$ for a village market to about
$10^9$ in a global market). Still such many-body dynamical
systems are statistical in nature and statistical physical
principles should be applicable. Remember, each constituents
particle in a gas follows some  well-defined  equations of
motion (say, Newton's equation for classical gases or 
Schr\"{o}dinger's equation for quantum gases), yet for the collective
behavior of  gases (or liquids or solids) we need to average
over the `appropriate' statistics for their stochastic behavior
in such `many-body' systems and calculate the emerging
collective or thermodynamic properties of the entire system.
Motivation to go for the `appropriate' statistics to estimate
the collective behavior or response of the  society comes
therefore very naturally. In the case of human agents in a
society, the corresponding equations governing individual
behavior are much more difficult and still unknown and
unpredictable, yet many collective social behavior are
quite predictable; ask any traffic engineer or engineers
designing stadium evacuation in panic situation.\\

\noindent
AS: Which problem of economics did Saha and Srivastava try to analyze
using Maxwell-Boltzmann distribution or statistical mechanics of ideal
gas?

\noindent
BKC: As can be seen from the example they had put to
the readers, they indicated to the students the problem
of income and wealth inequalities (they assumed
Gamma-function-like income distribution in~\cite{saha1931treatise};
reproduced in~\cite{chakrabarti2018econophysics}). They suggested to them that the
‘entropy maximization’  principle along with conservation
of money (or wealth) across the market (with millions of
transactions between the agents, buyers and sellers)
must be at  work in such ‘many-body’ social or market
systems. This will result in the consequent and  inevitable
inequality (equal distributions being entropically  unstable
against stochastic fluctuations, leading to steady state
unequal distributions).\\

\noindent
AS: That is quite interesting. Can you  elaborate a
bit more and explain a bit of statistical physics
specifically for the classical  ideal gas?

\noindent
BKC: Let me try. One can present the derivation of
the energy distribution among the constituent (Newtonian)
particles of a (classical) ideal gas in equilibrium at a
temperature $T$ as follows: If $n(\epsilon)$ represents
the number density of particles (atoms or molecules of
a gas) having energy $\epsilon$, then one can write
$n(\epsilon)d\epsilon = g(\epsilon)f(\epsilon, T)d\epsilon$.
Here $g(\epsilon)$ denotes the `density of
states' giving $g(\epsilon)d\epsilon$ as the number of dynamical states possible
for any of the free particles of the gas, having kinetic
energy between $\epsilon$  and $\epsilon + d\epsilon$
(as counted  by the different momentum vectors $\vec p$
corresponding to the same kinetic energy: $\epsilon =$
$|p|^2 /2m$, where $m$ denotes the mass of the particles).

Since the momentum $\vec p$ is a three dimensional
vector, $g(\epsilon)d\epsilon \sim |p|^2 d|p| \sim (\sqrt \epsilon) d\epsilon$. This is
obtained purely from mechanics. For completely stochastic
(ergodic) many-body dynamics or energy exchanges,
maintaining the the energy conservation, the energy
distribution function $f(\epsilon, T)$ should satisfy
$f(\epsilon_1)f(\epsilon_2) = f(\epsilon_1 + \epsilon_2)$
for any arbitrary choices of $\epsilon_1$ and $\epsilon_2$.
This suggests $f(\epsilon) \sim \exp(-\epsilon/\Delta)$,
where the factor $\Delta$ can be identified from the
equation of state for the gas (positive sign in the
exponential is neglected because of the observation that
the number decreases with increasing energy).
This gives $n(\epsilon) = g(\epsilon)f(\epsilon) \sim$
$(\sqrt \epsilon) \exp(-\epsilon/KT)$. Knowing this
$n(\epsilon)$, one can estimate the average pressure $P$ 
the gas exerts on the walls of the container having volume
$V$ at equilibrium temperature $T$ and compare with the
ideal gas equation of state $PV = NKT$ ($K$
denoting the Boltzmann constant). The gas pressure can
be estimated from the average rate of momentum transfer
by the atoms on the container wall and one can compare
with that obtained from the above-mentioned equation of
state and identify different quantities; in particular,
one identifies $\Delta = KT$.\\

\noindent
AS: How does one then extend this to the markets?

\noindent
BKC: Yes, let us consider the trading markets, where there
is no production (growth) or decay. Also, the total amount
of money (considered equivalent to energy) and number of
traders (or agents, considered as particles or `social
atoms') remain fixed or constant throughout the trades.
Since in the market money remains conserved as no one can
print money or destroy money (will end-up in jail in both
cases) and the exchange of money in the market is
completely  random, one would again expect, for any
buyer-seller transaction in the market, $f(m_1)f(m_2)$
$ = f(m_1 + m_2)$, where $f(m)$ denotes the equilibrium
or steady state distribution of money $m$ among the
traders in the market. This then, in a similar way,
suggests $f(m) \sim \exp(- m/\Delta')$, where $\Delta'$
is a constant. Since there can not be any equivalent of
the particle momentum vector for the agents in the market,
the density of states $g(m)$ here is a constant (any
real-number value of $m$ corresponds to one market state).
Hence, the number $n(m)$ of traders or agents having money
$m$ will be given by $n(m) = c \exp(- m/\Delta')$, where $c$
is a constant. One must also have $\int_0^M n(m)dm$
$ = N$, the total number of traders in the market, and
$\int_0^M mn(m)dm = M$, the total amount  money in
circulation in the market (or country). This gives, the effective `temperature' of the economy 
$\Delta' = M/N$, the average available money per
trader or agent in this closed-economy (as no growth,
migration of laborers, etc. are considered). This gives
exponentially decaying (or Gibbs-like) distribution of
money in the market (unlike the Maxwell-Boltzmann or
Gamma distribution of energy in the ideal gas), where most
of the people become pauper ($n(m)$ is maximum at $m = 0$).\\

\noindent
AS: Is this exponentially decaying income or wealth
distribution realistic for any economy?

\noindent
BKC: That  discussion  will take us to the  recent studies
by econophysicists and data comparisons. We will defer
those to the  next section (section 3). Indeed, some
success of the model (sketched above) in capturing the real
data has been explored extensively by Victor Yakovenko and his
group from Maryland University. We, in Kolkata, explored what
could make the distribution more like a  Gamma distribution,
as Saha and Srivastava indicated in their book~\cite{chakrabarti2018econophysics} to be
an observed  phenomenon. We also tried to capture the Pareto
tail of such a  distribution. Avoiding detailed discussion here,
we only refer here to three popular papers~\cite{dragulescu2000statistical,chakraborti2000statistical,chatterjee2004pareto} 
in this context. The model sketched above essentially follows~\cite{chakrabarti2018econophysics,dragulescu2000statistical}. In this model, the exchanged money or wealth
in each trade (equivalent to any of the particle-particle
collision  in Ideal gas) is completely random, subject to
conservation of money (or wealth). A trader, acquiring a
lot in earlier trades may lose the entire money or wealth
in the next trade as the total money (wealth) will be
conserved if the partner trader gets that. If one introduces
a saving propensity of each trader, so that each trader saves
a fraction of their individual money (wealth) before the
trade and exchanges randomly the respective rest amount
in the trade (keeping  total money or wealth again conserved)
the resulting steady state distributions capture the above
mentioned desirable features. One can easily see that,
unlike in  the Kinetic-exchange model described above,
the possibility for any trader (with non-vanishing saving
propensity) to become an absolute popper vanishes, as that
will require that trader to lose in every trade.
Consequently, the exponential distribution becomes unstable with effect to any non-vanishing saving propensity 
and the stable distribution will  become Gamma-like for
uniform saving  propensity of the traders~\cite{chakraborti2000statistical} and
initially Gamma-like but crossing over to Pareto-like
power-law decay when traders have non-uniform saving
propensities~\cite{chatterjee2004pareto}. These results are non-perturbative
results; any non-vanishing saving propensity will
induce these features;  the saving propensity magnitudes
only determine the most-probable  income (wealth) or the
income (wealth) crossover point for Pareto  tail of the
distribution.\\

\noindent
AS: Can we come back to your journey towards
econophysics? Apart from Saha-Srivastava's book, any
influence from other books, especially from economics?

\noindent
BKC: After Graduation and Post-Graduation from Calcutta
University, I joined, in  early 1975, the Saha Institute
of Nuclear Physics as a Research Fellow in Condensed
Matter Statistical  Physics for my Ph. D. degree.
By that time I had a large personal collection of
(mostly cheap editions, reprinted in India), general books, text
books, other books and monographs in subjects outside
physics; primarily in philosophy and economics. I
had attempted closer studies of some them  including: 
\textit{The Problems of Philosophy}, Bertrand Russell
(Cambridge Univ.), Oxford Univ. Press, Oxford (1959);
\textit{Mathematical Logic \& the Foundations of Mathematics:
An Introductory Survey}, Geoffrey Thomas Kneebone (Univ.
London), D. van Nostrand  Co. Ltd., London (1963);
\textit{The Problems of Philosophy}, Satischandra Chatterjee
(Univ. Calcutta), Calcutta Univ. Press, Kolkata (1964);
\textit{The Philosophy of Wittgenstein},  George Pitcher
(Princeton Univ.), Prentice-Hall Inc., New Delhi (1964);
\textit{An Introduction  to Philosophical Analysis}, John
Hospers (Univ. Southern California), Prentice-Hall
Inc., New Delhi (1971); Economics, Paul A. Samuelson
(MIT), Tata-McGraw Hill, New Delhi (1971); and \textit{ Economic Theory \&
Operations Analysis}, William J. Baumol (Princeton Univ.),
Prentice-Hall Inc., New  Delhi (1978).

\begin{figure}[H]
\begin{center}
\includegraphics[width=1.0\linewidth,height=0.7\textheight]{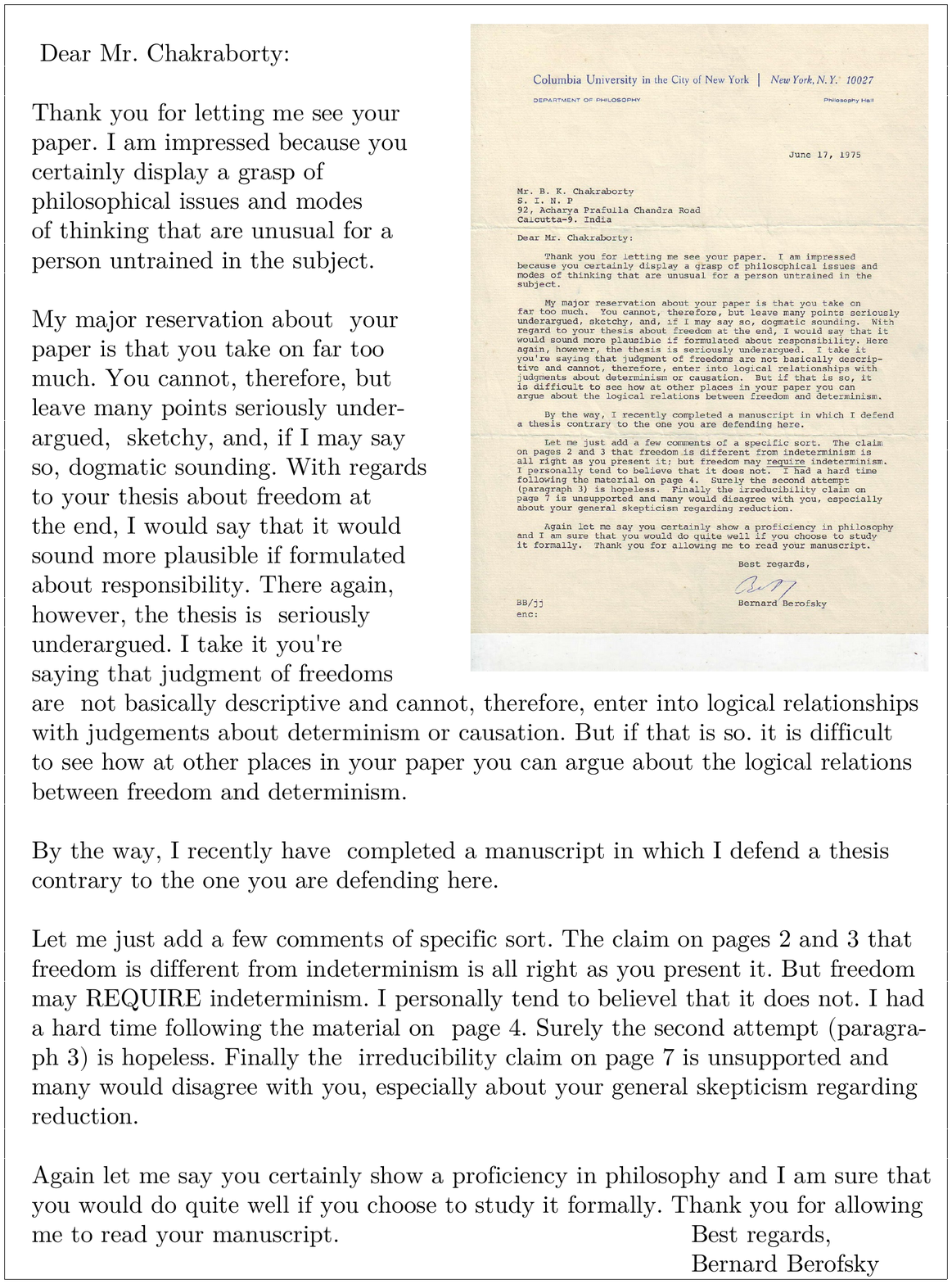}
\end{center}
\caption{ 
Reply (dated June 17, 1975) from Bernard Berofsky of the Philosophy
Department of Columbia University to BKC on his criticisms of Bernard’s
paper on `Indeterminism and Freedom'.
}
\label{fig_secII_I}
\end{figure}

I tried to go through some of the isolated chapters
or sections of these books, which I could understand, enjoyed or liked
most. Occasionally I got excited and tried my own
analysis, following them, on some interesting problems
or discussions coming in my way. One such piece was a
paper on `Indeterminism and Freedom' by Bernard
Berofsky of Columbia University, published in 1975,
perhaps in Philosophical Quarterly. Among others, it
also alluded to quantum physics in defending his thesis
on freedom. I wrote a note detailing my criticisms
and posted that to the  author. The  author, from
the Department of Philosophy, Philosophy Hall,
Columbia University in the City of  New York, wrote to me the following letter on June 17, 1975 (see Fig.~\ref{fig_secII_I}):\\

\noindent
AS: Obviously, you  did not follow his suggestion,
in fact, cordial invitation,  to switch over to
Philosophy. Why did you not?

\noindent
BKC: Though I was seriously thinking of switching over
to philosophy in a formal way, following Bernard's
suggestion, some quick apparent success in my physics
research publications  with the newly developed
Renormalization Group theory in those days kind
of blinded me and left me with two minds. Somehow,
I failed to take a decision and continued with my physics
research until I practically forgot about the  other
choice! In late 1978, I submitted my Ph. D. thesis in
Condensed Matter Physics to the University of
Calcutta and got the degree in 1979, and by the
end of that year, I  left for  post-doctoral researches
in the Theoretical Physics Department of the University
of Oxford and the Institute of Theoretical Physics,
University of Cologne.

I came back and joined the Saha Institute of Nuclear
Physics as Lecturer in 1983, and I started my
research in statistical physics with four Ph. D.
students joining me simultaneously (including
Subhrangshu Sekhar Manna, who later developed the
`Manna Model', belonging to the `Manna Universality
Class'). Soon the statistical physics research in our
group became  so engaging and happening (with sixteen
Ph. D. students, so far, getting their Ph. D. degrees
and several of them becoming quite well known later
for their pioneering researches and still collaborating
with me), I did not get much time until early nineties
when I decided to try some research on `economics-inspired physics'. I went back to the problem Saha and
Srivastava addressed in their textbook mentioned above
and I co-organized a conference in January 1995, together
with some  established Indian  statistical physicists and
(reluctant!) economists as participants. In the
Proceedings of the Conference, I published (together
with an economist Sugata Marjit), my first paper~\cite{chakrabarti1995self}
dealing with statistical physics of Income distribution
and related problems.

\begin{figure}[H]
\begin{center}
\includegraphics[width=1.0\linewidth,height=0.45\textheight]{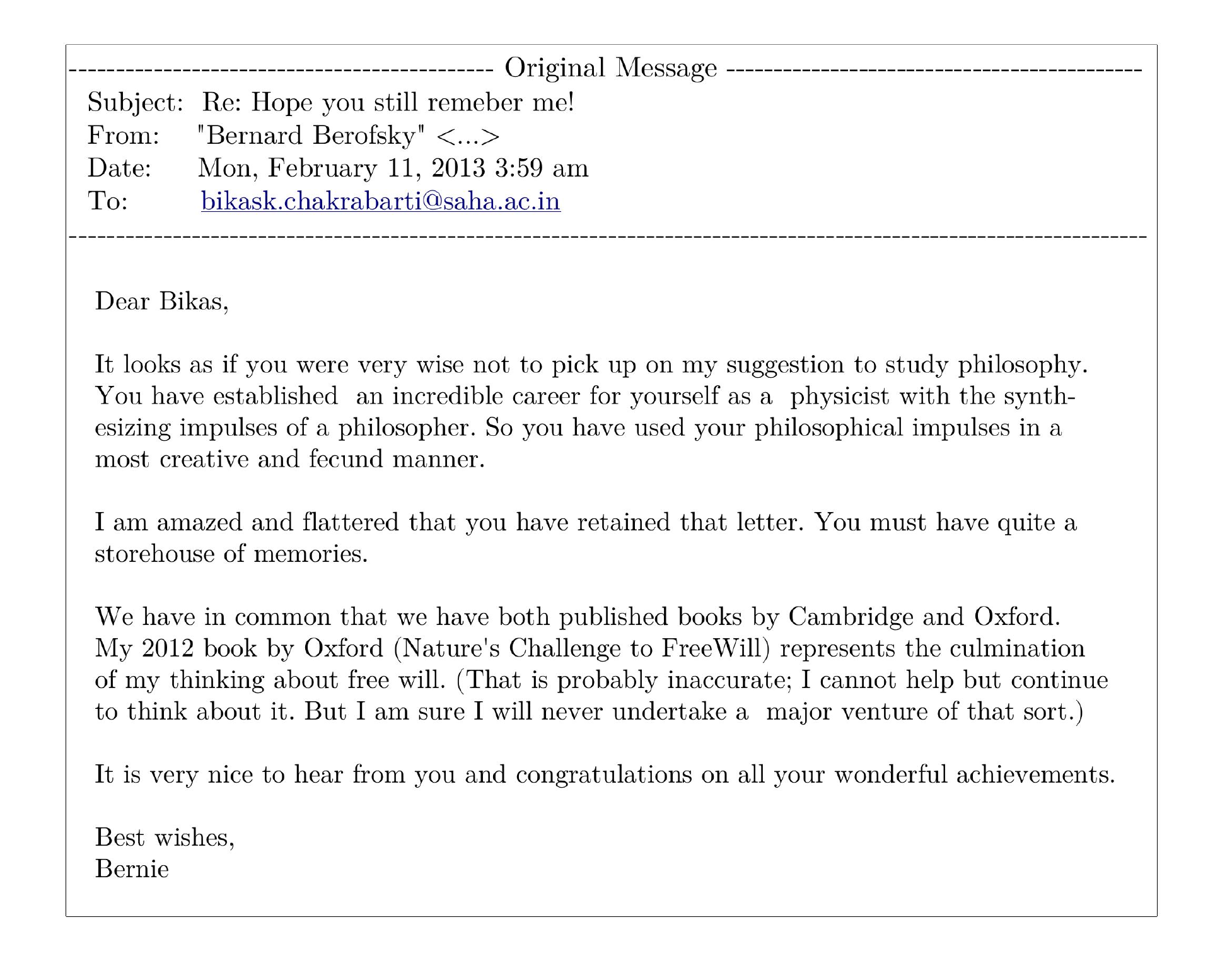}
\end{center}
\caption{Mail from Bernard Berofsky of the Philosophy  Department of Columbia
University, in response to BKC's surprise contact mail in 2013 (after almost
thirty seven years!),  appreciating and identifying the development of
econophysics as one due to the ``physicist(s) with synthesizing impulses of
a philosopher ... (using) philosophical impulses in a most creative and
fecund manner''.}
\label{fig_secII_II}
\end{figure}

By the end of the year, as a part
of the StatPhys-Kolkata II (series of International
Conferences organized by us in Kolkata every 3-4 years,
latest event StatPhys-Kolkata X, held end of 2019), we
had organized a special session on `Economics-Inspired
Physics' and Eugene Stanley in his talk coined the term
'Econophysics' and had put that in the title of his paper~\cite{stanley1996anomalous}
 published in the Proceedings of the conference in
Physica A, vol 224 (1996).

Though econophysics was quite
risky as a topic of Ph. D. research in late nineties (even
today; still no faculty position in econophysics in our
country, or for that matter, hardly exists  elsewhere in
the world), two brave students (Anirban Chakraborti and
Arnab Chatterjee) expressed forcefully their desire  to
join the research on eventual topic of `econophysics'. I was
also fortunate, my colleague Sitabhra Sinha also joined us
in such investigations. In the last 25 years, since that
conference, significant developments have taken place in
the subject, and many of them will be covered
this special issue of Entropy.\\

\noindent
AS: We will come back to those developments later. I
understand, most of the papers on econophysics researches
are published in physics journals and not in economics
journals. What is the cultural level of appreciation
by the intellectuals today?

\noindent
BKC: This is indeed very difficult to answer. To tell
very  frankly, the response so far is not very supportive
or  encouraging!  Although, it must be mentioned, the
term econophysics has now entered in dictionaries
of economics (see e.g.,\cite{rosser2008econophysics}) and Encyclopedias
of social science and philosophy (see e.g.,~\cite{kaldis2013encyclopedia}).
That brings me to an interesting, rather recent,
correspondence  with my old philosopher `guide'
Bernard Berofsky in January 2013,  after thirty
seven years! This was quite accidental, when I
came across in my internet search a new book
published by him. I contacted him (giving a link
to my homepage) saying, ``sorry, I could not follow
your advice so far and had been very shy to
contact you. Now that I have become sixty, I
acquired sufficient  courage and ...". Bernard
immediately responded (see Fig.~\ref{fig_secII_II}) praising the development of econophysics due to the philosophical impulses of physicists.\\

\noindent
AS: Do you see really a philosophy behind econophysics?

\noindent
BKC: Yes, indeed. I wrote about it earlier also (see e.g.,~\cite{kaldis2013encyclopedia,chakrabarti2016can}]). I am not aware of all the documents on the
mutual connection between philosophy and econophysics.
I mentioned earlier about the entry on Econophysics in the
Encyclopedia of Philosophy and Social Sciences~\cite{kaldis2013encyclopedia}. I
came to know of a rather recent entry on Social Ontology in
The Stanford Encyclopedia of Philosophy~\cite{epstein2018social} which, in
the context of `social atomism', writes ``The idea is to
model societies as large aggregates of people, much as
liquids and gases are aggregates of molecules, ...''. Then,
after introducing the readers to two historical examples
of Quetelet in 1848 [Adolphe Quetelet, 1848, Du système
social et des lois qui le régissent, Paris: Guillaumin]
and of Spencer in 1895 [Herbert Spencer,  1895, The
Principles of Sociology, New York: Appleton] it says
``Contemporary representatives include models in
sociophysics and econophysics (see Chakrabarti et al.
2007) ... [which] take a society or market to be an
aggregate of these interacting individuals [Bikas K.
Chakrabarti, Anirban Chakraborti and Arnab Chatterjee, 2007,
Econophysics and Sociophysics: Trends and Perspectives,
Hoboken, NJ: Wiley]''.

Let me now go back to the main discussion and reiterate my basic argument in favor
of considering economics as a natural science. Our knowledge
about truth can, epistemologically speaking, be either
deductive or inductive. Mathematics is an usual example
of the deductive knowledge (though not all of it can be
deduced from axiomatic logic); mathematical truths do not
require any laboratory test or `observational' support
from the `nature' to prove or validate them. Linguistically
speaking, it is like the tautology ``A bachelor does not
have wife''. One does not need to check each and every
bachelor to confirm truth of the statement – first part of
the sentence confirms the second part. The same is true
about the statement ``two plus two equals four''.
Mathematical truths are analytical truths; left hand side
equals (in every intention   and content)  the right
hand side. Mathematics therefore is not directly a natural
science~\cite{whitehead1913principia}; though it has been at the root of the logical
structure of many natural sciences; physics in particular.
Natural sciences, however, are basically inductive in nature.
They are based on natural or (controlled) laboratory
observations. The statement “The sun rises every twenty
four hours on the east in the morning” is not a tautology
nor an analytical truth. Though east may be defined as the
direction and morning may be defined by the time of sunrise,
that it rises every twenty four hours, is an inductive
(or empirically observed) truth, and therefore tentative
(and not like mathematical truths which are analytical and
certain). Natural sciences start with observations and
end in observations; both using inductive reasoning or
logic; in-between they often  employ the tools of deductive
logic, mathematics (as most condensed form of deductive
logic).\\

\noindent
AS: So,  the tools of mathematics and logic are employed to
find and establish relationships among these `natural'
observations to develop natural sciences. Where does then
economics belong to?

\noindent
BKC: That is the crucial question. Intermediate analysis
using mathematics is just applied mathematics, and can
not be considered as (pure) mathematics. Any branch of
natural science does that. Economics has been and will
be a part of natural science, where natural observations,
not much of controlled or laboratory observations, need
to be analyzed employing deductive  logic and mathematics.
Economics therefore should  naturally belong to natural
sciences!\\

\noindent
AS: Agreed. But why econophysics?

\noindent
BKC: You see, in natural sciences today, there
are several branches or disciplines like physics,
chemistry, biology, geology, etc.. The
differences are not natural and certainly nature
did not create them: they are human creations.
The demarcations among these disciplines are not
always clear. As we mentioned earlier, there are
clear differences (in the nature of logic
employed) between mathematics and natural sciences.
But that does not extend to the branches of
natural sciences. In a white light spectrum, our
color perception continuously change from violet
to red (without any sharp boundary) as the
wavelength changes in this collection of
electromagnetic waves. Similar are the cases of
the different branches of natural science.
They are not strictly differentiable; are
historical in  origin and continued by us for
our own convenience (during upbringing; like
perhaps religion; both are man-made). Of course,
it is hard today to be an expert in the whole of
even one branch of natural science. We therefore
try to learn and acquire expertise in one
sub-branch or a sub-sub-branch of natural science.
An unique feature of the sub- or sub-sub-branches
of natural science is that an established `truth'
or a 'fundamental law' in one branch does not
become 'false' or 'wrong' in another; only
importance varies from discipline to discipline;
quantum physics or gravity laws do not become
invalid or wrong in chemistry or biology or
mineralogy. Only gravity laws may be less important
in chemistry or biology or mineralogy and
vice-versa. Models of geomagnetism in earth science
can not be built upon a law contradicting Maxwell's
laws of electromagnetism. Developments in younger
branches in science therefore profitably utilized
earlier established laws or ideas in older
branches of natural science.  Many of the early
successful scientists (even some mathematicians)
happen to have been identified  as physicists and,
consequently, physics has become like an `elder
brother' among natural sciences, and it is now
equipped with a huge armory of ideas, laws and
models to comprehend the nature. Economics as a
relatively newer entrant to natural science can
therefore expect gainful advantages from such
econophysical attempts!\\

\noindent
AS: I remember you once told me that the concept of modeling
dynamics of physical systems and of economics systems are
fundamentally different. Can you elaborate the point in
this context?

\noindent
BKC: I do not remember which point we had been discussing.
However, there is a typical one which may be discussed.
Modeling dynamics of a physical system like a particle,
using, say, the Newton's equation of motion, gives its
dynamical state at a later time $t$ by solving the equation
of motion and utilizing the information  regarding its
dynamical state at an earlier time (say, at $t = 0$;
called initial conditions). Exact solutions may not be
possible as in the thermodynamic or many-body systems, but
based on the statistical characterization  of the state
of the system at an earlier time, the dynamical formulation
helps solving the statistics of the system for any future
time. The economic agents or organizations, even under
nominally identical economic situation, may have
(continually upgradeable) anticipation about the future
and the model dynamics need to accommodate, along with
their initial economic state,  such anticipatory factors
(which are continually adjusted or learned through the ongoing 
dynamics itself!) to solve for the future. Such
self-consistent `learning' dynamics of physical systems
are not typical, though some recent many-body game
theoretic models, with iterative learning for optimal
use of scarce resources as in the binary-choice Minority
Game (see e.g.,~\cite{challet2013minority}), or many-choice Kolkata Paise
Restaurant Problem (see e.g.,~\cite{chakrabarti2017econophysics}) naturally
incorporate such evolving learning features in the
self-correcting dynamics themselves. We will discuss
some details of the later problem here. In any case,
these studies are new and still very limited in scope.\\

\noindent
AS: To summarize, though many of you had started your
econophysics researches more than twenty five years
back, since Gene Stanley coined the term econophysics
in 1996 (in his publication~\cite{stanley1996anomalous} in the Proceedings
of the second StatPhys-Kolkata Conference), and many
more physicists joined after that, the subject is not
established yet.

\noindent
BKC: You are partly right. In fact, physicists
have long been trying to formulate and comprehend
various problems of economics. As mentioned
before, since 1931, the statistical physics
modelling of income and wealth distributions are
being tried. However, these older physics
attempts had been sporadic and isolated ones;
physicists, successful in such attempts, like Jan
Tinbergen (Economics Nobel Prize winner in 1969;
had Ph.D. in statistical Physics under Paul
Ehrenfest of Leiden University) had to migrate
to economics department. Since 1996 (more
correctly perhaps since 1991, when Rosario
Mantegna published his paper~\cite{mantegna1991levy} on Milan stock
exchange data modelling), however, the situation
has changed considerably. Physicists are now
investigating economics problems along with their
students and colleagues from the same department
and are publishing their econophysical research
papers in physics journals (in around 2000,
Econophysics had been assigned the Physics and
Astronomy Classification Scheme (PACS) number
89.65Gh by the American Institute of Physics).

   I personally think, however, that an
intensive and successful branch of
econophysics research started with Scott
Kirkpatrick and coworkers in 1983 when they
proposed~\cite{kirkpatrick1983optimization} the idea and technique of
`Simulated Annealing' (or `Classical Annealing')
to get practical solutions of the computationally
hard multi-variable optimization problems like the
(managerial) economics problem of the Traveling
Salesman Problem (TSP), using tuning (annealing)
of Boltzmann-type fluctuations (simulating
thermal ones) to escape from the local minima
to reach eventually one of the (degenerate)
global minima of the cost function (travel
distance). This is a very successful story of
the application of (statistical) physics to solve
a problem which in nature and basic intent a
(financial) economics problem involving
multi-variable optimization. It may be noted in
this connection that the technique has since been
applied to all branches of science as well as
technology and the original paper~\cite{kirkpatrick1983optimization} has received major
attention of scientists and engineers (so far
received more than 48,000 citations, according to
google scholar). This idea still continues leading
to a very intriguing and active domain of research
in computationally hard problems of optimizations,
using statistical physics and physics of spin
glasses. This eventually led to the concept and
technique of `Quantum Annealing' (or of `Stochastic
Quantum Computing'), where simulated quantum
fluctuations (instead of simulated thermal
fluctuations) are profitably used to tunnel
through high but narrow local barriers~\cite{das2008colloquium},
separating the global minima or solutions (see
e.g.,~\cite{santoro2006optimization} for a brief review on solving TSP
using quantum annealing). As I discussed earlier
in my Econophysics-Kolkata Story~\cite{chatterjee2005econophys}, we started
in 1986 (see subsection 3.A) investigations on
the statistical physics of the TSP. Soon my student
Parangama Sen joined the effort~\cite{sen1989travelling}. (She
eventually concentrated more on Sociophysics and
developed, among others, the  Biswas-Chatterjee-Sen
model, see e.g.,~\cite{sen2014sociophysics}, for collective opinion
formation together with our students Soumyajyoti
Biswas and Arnab Chatterjee.  In this connection,
let me take the opportunity to acknowledge the
contributions  of my other students, Srutarshi
Pradhan, Asim Ghosh and Sudip Mukherjee,
Suchismita Banerjee, and of course you Antika, and
of my colleagues in the Kolkata-econophysics group,
namely Anindya Sundar Chakrabarti, Manipushpak
Mitra and Satya Ranjan Chakravarty, allowing us to
make some significant contributions to
econophysics, which we are going to summarize in
the next section).\\

\noindent
AS: So, you think that successful researches in
econophysics already started with the Simulated
Annealing paper by Kirkpatrick et al. in 1983,
although econophysics researches on more popular
economics problems started in 1990’s and, more
specifically, after Stanley coined the term in
1996?

\noindent
BKC: Yes, you are right. We will discuss in little more
details (in the next section; subsection 3.A) the
impact of statistical physics in developing the
Simulated (Classical or Quantum) Annealing techniques
for the financial computation problems involving
multi-variable optimization  of the Traveling Salesman
type. The inspiring  success of the classical annealing
technique, initiated  by the Simulated Annealing method,
has led to several intriguing developments in statistical
physics and to many applications in computer science.
Further potential extension  in the context of solving
NP-hard problems using quantum annealing has become
one of the core research topic today in quantum many-body
(statistical) physics and in quantum computation. Indeed
I consider this outstanding development of simulated
(classical or quantum) annealing techniques (starting
with Kirkpatrick et al.~\cite{kirkpatrick1983optimization}, see also~\cite{das2008colloquium}) for the
Traveling Salesman type multi-variable optimization
problems to be a landmark achievement in the true spirit of
econophysics. Of course the present phase of
econophysics research activities stemmed from
several influential papers, analyzing financial
market fluctuations, by Rosario Mantegna and
Eugene Stanley and in particular following the
publication of Kolkata Conference Proceedings
paper~\cite{stanley1996anomalous} by Stanley et al. in 1996.\\

\section{Major Achievements and Publications of the `Kolkata School'}
Physicist Victor Yakovenko and economist J.
Barkley Rosser in their pioneering interdisciplinary
collaborative review~\cite{yakovenko2009colloquium} in the Reviews of
Modern Physics (2009) on  econophysics of
income and wealth distributions, discussed
about some of the `influential' and `elegant'
papers from the `Kolkata School'. We will
briefly summarize in this section some of
our major researches in econophysics
(including those on wealth distributions).\\

\noindent
A) Traveling Salesman Problem and Simulated (Classical
\& Quantum) Annealing

As already  discussed, the Traveling Salesman Problem or
TSP is, in its intent and structure, a computationally
involved financial management problem (see e.g.,~\cite{orman2006survey,rasmussen2011tsp}). 
The problem can be easily
defined as a geometric one. Suppose in an unit square area
 there are $N$ random dots, representing the
cities. The salesman has to make a visit to all
the cities and come back with minimum travel cost.
The travel cost to visit all these cities  will
depend on the total travel distance of the tour. Each
component of the travel distance  between any two
cities can be taken as the Euclidean distance (or
as appropriate for the spatial metric, say Cartesian) between them.
One can easily check that there are $N!/2$ (growing
faster than exponential in $N$) distinct tours or
trips to visit all the $N$ cities. Obviously, all of
these trips do not have identical value (`cost') for
the total travel length ($D$), and the problem is
to find the trips(s) which will correspond to the
minimum travel distance $D$. Searching over all the
possible trips soon  becomes impossible as $N$
becomes large, and  there is no perturbative way to
improve on any randomly chosen travel path to reach the
global solution. At any point or city on a tour, there
are $N$ order choices for the next move or visit and
the optimization problem of the total travel distance
is truly a multi-variable one. It may be noted that the
problem becomes trivial in one dimension (homes or
offices placed randomly on a straight road), where
the salesman can start from one end of the road and
finish at the other end). Generally, for two dimension onwards, search time for such a minimum `cost' (from among $\exp(N)$ number of
trips or configurations), can not be
bounded by any (deterministic) polynomial in $N$ (NP-hard problem).

From now onwards, let us concentrate our discussion on TSP in two dimension. 
The scale of the total travel distance, however, can be
easily guessed using a `mean field' picture. If $N$
randomly placed points (cities) fill an unit (normalized) area,
then  the `average' or `mean' area  per city is $1/N$,
giving nearest neighbor distance to be of order
$1/\sqrt N$ and total travel distance $D =
\Omega \sqrt N$. Numerical estimates suggests
$\Omega \simeq 0.71$~\cite{percus1996finite}.

The problem is truly global in nature. Choice of the
next city to visit depends on the position of even the
farthest city in the country. However, one can
approximately solve the problem (see~\cite{sinha2010econophysics}) by
reducing it to an effective one dimensional problem
where the country (unit square) is divided into
hypothetical parallel strips of width $w$ and
the salesman visits the cities within each strip in a
`directed' way and the total travel distance $D$ is
optimized with respects to single variable $w$ (optimal
value then grows as $\sqrt N$) and gives (see e.g.,~\cite{beardwood1959shortest,sinha2010econophysics}) 
$\Omega \simeq 0.92$.  Another  way is to put
the cities randomly with concentration  $\rho$ on  the
lattice sites of, say, an unit square lattice. The
lattice constraint can help then the calculation of
the optimal  travel distance. The optimal (normalized)
travel path length then  scales as $D = \Omega \sqrt \rho$.
At $\rho = 1$, the lattice constraints would immediately
imply that the global search problem  reduce to a
local one and all the space filling Hamiltonian walks
would correspond to optimized tour with $\Omega = 1$.
In the approximate single variable solution (minimization of $D$ with respect to $w$) indicated above, the strip width $w$ grows as $1/\sqrt \rho$ as $\rho$ decreases.
For $\rho \to 0$,  however, the lattice constraints
disappear and  the problem reduces to TSP on
continuum as defined earlier (NP-hard, $w \to \infty$, with
$ \Omega \simeq 0.71$~\cite{percus1996finite}). Where does the problem become 
NP-hard? This study was initiated by us (see~\cite{sen1989travelling,chakrabarti1986directed,dhar1987travelling,ghosh1988travelling,chakraborti2000travelling})
and they indicated (see also~\cite{sinha2010econophysics}) that the TSP on dilute lattices becomes NP-hard only at $\rho \to 0$ 
(though this is not settled yet and some arguments support that it crosses over to NP-hardness at $\rho = 1_-$ or  
as soon  as $\rho$ becomes less than unity).

As already mentioned earlier (in section 2) a major
computational breakthrough of TSP and other
such multi-variable optimization problems came from
the 1983 seminal paper on Simulated Annealing’~\cite{kirkpatrick1983optimization} 
by Kirkpatrick et al., who proposed a novel
stochastic technique, inspired by the metallurgical
annealing process and statistical physics of
frustrated systems.

Imagine a bowl on the table and you need to `locate'
its  bottom point. Of course, one can calculate the
local depths (from a reference height) everywhere
along the inner volume of the bowl and search for the
point where the local depth is maximum. However, as
every one would easily guess, a much simpler and
practical method would be to allow rolling of a marble
ball along the inner surface of the bowl and wait for
locating its resting position. Here, the physics of the
forces of gravity and friction allows us to `calculate'
the location of the bottom point in an analog way!
Algorithm wise, it is simple. For any possible  move,
if the changed `cost' function has lower value, one
should accept the move and reject it otherwise. Success
for the search of the minimum is guaranteed. In
principle, a similar trick would work for cases where
the bowl becomes larger and its internal surface gets
modulated, as long as the surface contour or `landscape'
has valleys all tilted towards the same bottom point
location. Computationally hard problems arise when
these valleys are separated by `barriers', which are
(macroscopically) high. The simulated annealing
suggests a way out to overcome (at least for finite
height  barriers) by allowing moves costing higher
to have (Gibbs-like) lower probability of acceptance.

To search for the optimized cost (travel distance in
TSP  or energy of the ground state(s) in spin glasses) at
eventually vanishing level of noise (or `simulated
temperature'), one starts  from a high noise
(temperature) `melt' phase, and tune slowly the noise
level. In this `simulated' process, the (classical)
noise at any intermediate level of annealing allows for
the acceptance of the changed `costs' $\Delta D$ in
distance or energy $D$: 100\% acceptance of the move if
$\Delta D < 0$ and acceptance of the move with a
Gibbs-like probability $\sim \exp (-\Delta D/T)$ for
moves with increased in cost $(\Delta D > 0)$). As the
noise level  ($T$) is slowly reduced during the
annealing process,  the gradually  decreasing
probability  of accepting higher cost values, allows
the system to come out of the local minima valleys
and settle eventually in (one of)  the `ground state'
(with minimum $D$) of the  system. For slow enough
decrease of noise $T(t)$ with time $t$, one can
estimate the quasi-equilibrium (thermal) average
of the cost function $<D>$ at ant time $t$ and
derive the effective `specific heat' value
$\delta <D>/\delta T$ as a function of $t$. One needs to
be very slow ($|dT/dt|$ very small) when the effective
specific heat increases with decreasing $T$,
indicating the `glass' transition point  and anneal
at faster rates on both sides of the transition point.

\begin{figure}[H]
\begin{center}
\includegraphics[width=7.5cm]{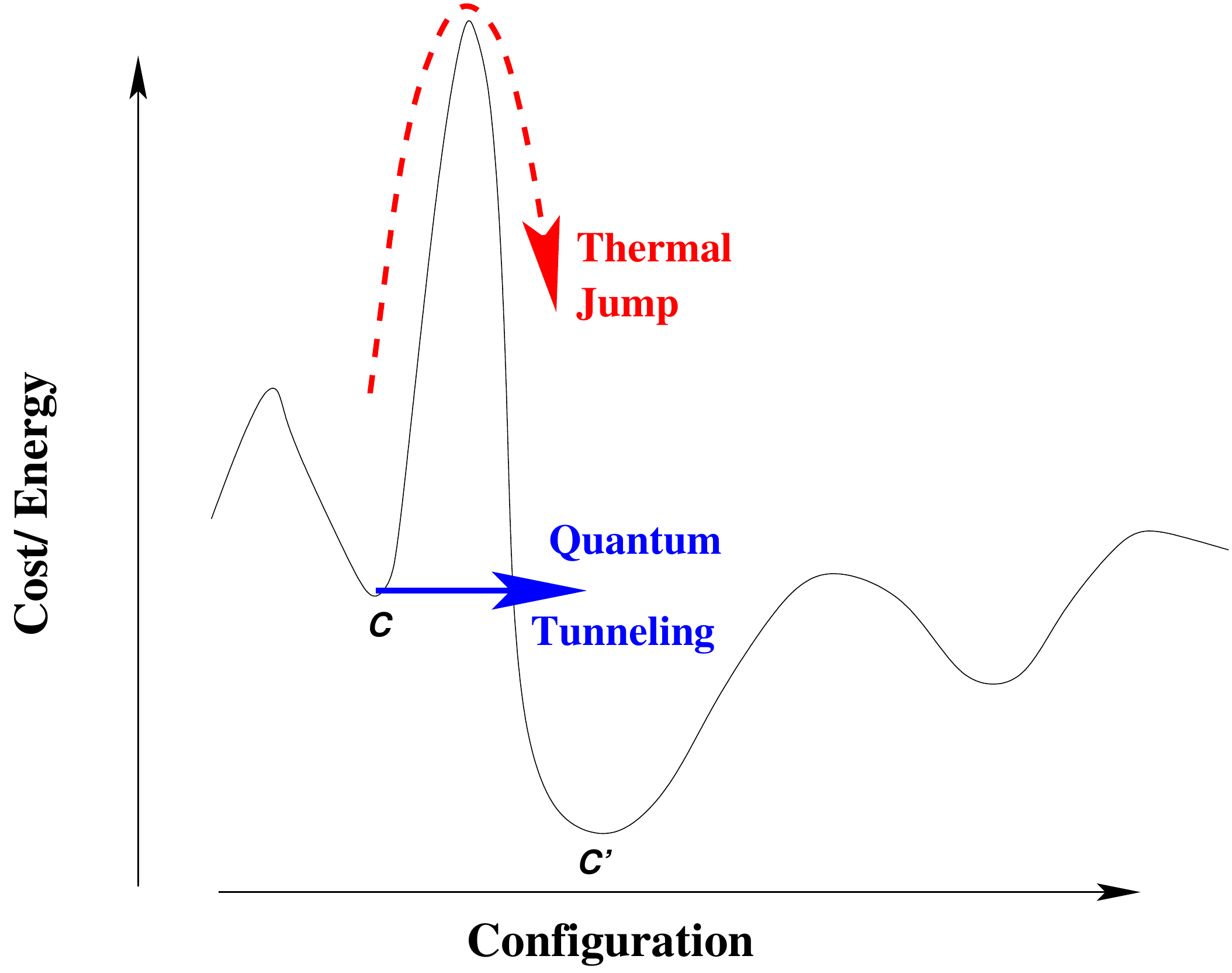}
\end{center}
\caption{While optimizing the cost function of a
computationally hard problem (like the minimum
travel distance for the TSP), one has to get out
of a shallower local minimum  like the
configuration C (travel route), to reach
a deeper minimum C'. This requires jumps or
tunneling like fluctuations in the dynamics.
Classically one has to jump over the energy or
the cost barriers separating them, while quantum
mechanically one can tunnel through the same. If
the barrier is high enough, thermal jump becomes
very difficult. However, if the barrier is narrow
enough, quantum tunneling often becomes quite easy.
Indeed, assuming the tall barrier to be of height
$N$ and width $\tilde{w}$, one can estimate (see e.g.,~\cite{mukherjee2015multivariable}), the tunneling probability through the
barrier to be of order $\exp[-(\tilde{w}\sqrt N)/\Gamma]$
where $\Gamma$ denotes the strength of quantum
fluctuations (instead of the the classical escape
probability of order $\exp [-N/T]$, $T$ denoting the
thermal or classical fluctuation strength). }
\label{fig_secIIIA_I}
\end{figure}

As has been indicated in the earlier section, it has
been a remarkably successful trick for `practical'
computational solutions of a large class of
multi-variable optimization problems, as in most
multi-city travel cost optimizations and similar
multi-variable optimizations (see e.g.,~\cite{orman2006survey,rasmussen2011tsp,bonomi1984n,zhou2019traveling}).

Though some `reasonable' optimization can be
achieved very quickly using appropriate annealing
schedules, the search time for reaching the lowest
cost state or configuration for NP-hard problems
however grows still as $\exp(N)$. The bottleneck
could be identified soon. Extensive study of the
dynamics of frustrated random systems like the
$N$ spin (two state Ising) glasses, in particular
of the Sherrington-Kirkpatrick variety (see e.g.,~\cite{das2008colloquium} 
also for a TSP version of the quantum
annealing), showed that its (free) energy landscape
(in the `glass' phase) is extremely rugged and
the barriers, separating the local valleys, often
become $N$ order implying the search for the
degenerate ground states from $2^N$ (or $N!/2$) states
is NP-hard (for the $N$-city TSP). In the macroscopic
size limit ($N$ approaching infinity) therefore such
systems effectively become non-ergodic or localized
and the classical (thermal) fluctuations like that in
the simulated annealing fail to help the system to
come out of such high barriers (at random locations
or configurations, not dictated by any symmetry)
as the escape probability is of order $\exp(-N/T)$ only.
Naturally, the annealing time (inversely proportional
to the escape probability), to get the ground state of
the $N$-spin Sherrington-Kirkpatrick model, can not be
bounded by any polynomial in $N$.

The idea proposed by Ray et al.,~\cite{ray1989sherrington} was that
quantum fluctuations in the Sherrington-Kirkpatrick
model can perhaps lead to some escape routes to
ergodicity or quantum fluctuation induced delocalization
(at least in the low temperature region of the spin glass
phase) by allowing tunneling through such macroscopically
tall but thin (free energy or cost functions) barriers
which are difficult to scale using classical fluctuations. 
This is based on the observation that escape probability
due to quantum tunneling, from a valley with single
barrier of  height $N$ and width $\tilde{w}$, scales as $\exp
(-\sqrt N \tilde{w}/\Gamma)$, where $\Gamma$ represents the
quantum (or tunneling) fluctuation strength (see Fig.~\ref{fig_secIIIA_I}). 
This extra handle through the barrier width
$\tilde{w}$ (absent in the classical escape probability of order
$\exp (-N/T)$) can help in a major way in its vanishing
limit. Indeed, for a single narrow ($\tilde{w} \to 0$) barrier
of height $N$, when $\Gamma$ is slowly tuned to zero, the
annealing time to reach the ground state or optimized cost,
will become $N$ independent (even in the $N \to \infty$
limit; $\delta$-function barriers are transparent to
quantum fluctuations, while classical or thermal
annealing to escape from such a barrier is impossible)!
It has led to some important clues. Of course,
complications  (localization) may still arise for many
such barriers  at random `locations'. In any case,
with this observation and some more developments,
the quantum annealing technique was finally launched
through the subsequent publications of a series of
landmark papers (both theoretical and experimental;
see~\cite{das2008colloquium}) and through a remarkable practical
realization of the quantum annealers by the D-wave
Group~\cite{johnson2011quantum}.

\begin{figure}[H]
\begin{center}
\includegraphics[width=8.5cm]{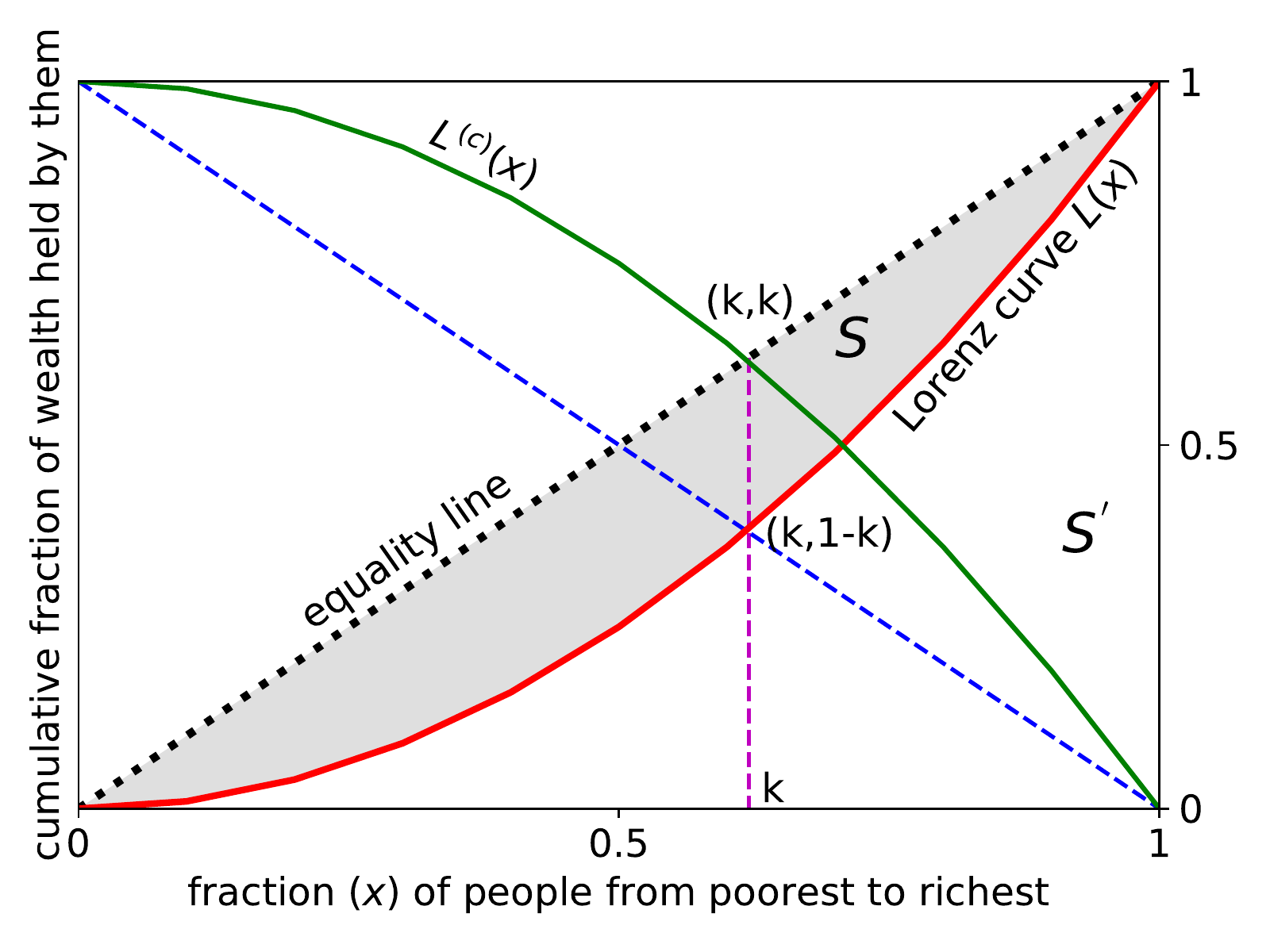}
\end{center}
\caption{ Lorenz curve (in red) or function $L(x)$
here represents the fraction of accumulated wealth against
the fraction $x$ of people possessing that, when
arranged from the poorest to the richest. The
diagonal from the origin represents the equality
line. The Gini index $(g)$ can be measured by
the area $(S)$ between the Lorenz curve and the
equality line (shaded region), normalized by the
total area $(S + S' = 1/2)$ under the equality line:
$ g = 2S$. The complementary Lorenz function $L^{(c)}(x) \equiv 1-L(x)$ is shown by the green line. The Kolkata index $(k)$ can be measured by the
ordinate value of  the intersecting point of the
Lorenz curve and the diagonal perpendicular to
the equality line. By construction, $L^{(c)}(k)$ = $1 - L(k) = k$,
saying that $k$ fraction of wealth is being possessed
by $(1 - k)$ fraction of richest population. }
\label{fig_secIIIB_I}
\end{figure}

Let us now conclude this subsection. 
Simulated Annealing technique, invented by
Kirkpatrick et al. in 1983~\cite{kirkpatrick1983optimization}, has since been
applied extensively also to solve problems of
collective  decision making in economics and social
sciences (see e.g.,~\cite{lucas2014ising} for a recent
review). As mentioned earlier~\cite{chatterjee2005econophys}, our group
started investigations on statistical physics
of TSP in 1986. The intriguing physics of Simulated
Annealing inspired us to explore the possible
further advantages of quantum tunneling (to allow
escape through macroscopically tall but thin
barriers in some NP-hard cases), where classical
annealing (using  thermal fluctuations) fails.
This led finally to the  quantum extension or to
the invention of the quantum  annealing technique,
where our initial contributions (~\cite{ray1989sherrington,das2008colloquium})
are considered to be important and pioneering.
See e.g.,~\cite{santoro2006optimization} for a brief review and~\cite{dong2020quantun} 
for some recent discussions on the advantages of
applying quantum annealing method to solve TSP.
Quantum annealing is a very active research field
today in quantum statistical physics and
computation (see e.g.,~\cite{tanaka2017quantum,albash2018adiabatic} for
recent reviews).\\

\noindent
B) Social Inequality Measure and Kolkata Index

Social inequality, income or wealth inequality in
particular, are ubiquitous. There  are several indices
or coefficients, used to measure them. The oldest and
most popular one being the Gini index~\cite{gini1921measurement}.

It is based on the Lorenz curve or function~\cite{lorenz1905methods} 
$L(x)$, giving the cumulative fraction of (total
accumulated) income or wealth possessed by the
fraction ($x$) of the population, when counted from
the poorest to the richest (see Fig.~\ref{fig_secIIIB_I}). If the income
(wealth) of every one would be identical, then $L(x)$
would be a straight line (diagonal)  passing through
the  origin. This diagonal is called the equality line.
The Gini coefficient ($g$) is given by  the area
between the Lorenz curve and the equality line
(normalized by the area under the equality line: $g = 0$
corresponds to equality and $g = 1$ corresponds to extreme
inequality.

We proposed~\cite{ghosh2014inequality} the Kolkata index or $k$-index given
by the ordinate value of  the intersecting point of the
Lorenz curve and the diagonal perpendicular to
the equality line (see also~\cite{ghosh2016inequality,chatterjee2017socio,sinha2019inequality,banerjee2020kolkata,banerjee2020social}). By construction, $1 - L(k) = k$,
saying that $k$ fraction of wealth is being possessed
by $(1 - k)$ fraction of the richest population. As
such, it gives a quantitative generalization of the
approximately established (phenomenological) 80-20 law
of Pareto~\cite{wiki_pareto}, saying that in any economy,
typically about 80\% wealth is possessed by only 20\% of the
richest population. Defining the complementary Lorenz
function $L^{(c)}(x) \equiv [1 - L(x)]$, one gets
$k$ as its (nontrivial) fixed point (while Lorenz
function $L(x)$ itself has trivial fixed points at
$x = $ 0 and 1). $k$-index can also be viewed as the
normalized $h$-index~\cite{hirsch2005index} for social inequality;
$h$-index is given by the fixed point value of the
nonlinear citation function against the number of
publications of individual researchers. 
We have studied the mathematical structure of
$k$-index in~\cite{banerjee2020kolkata} (see~\cite{banerjee2020social} for a recent
review) and its suitability, compared with the
Gini and other inequality indices or measures,
in the context of different social statistics,
in~\cite{ghosh2014inequality,ghosh2016inequality,chatterjee2017socio,sinha2019inequality}. 
See also~\cite{subramanian2015more}
 for redefining a generalized Gini
index and~\cite{sahasranaman2020spread} for a recent application
in characterizing the statistics of the
spreading dynamics of Covid-19 pandemic in
congested towns and slums of the developing
world.

In summary, inspired by the observations of richer
structure (self-similarity) of the (nonlinear)
Renormalization Group equations near the fixed
point (see e.g.,~\cite{fisher1998renormalization}), or of the nonlinear
chaos-driving maps near the fixed point (see
e.g.,~\cite{feigenbaum1983universal}) and noting that inequality
functions (such as the Lorenz function $L(x)$
or the Complementary Lorenz function $L^{(c)}(x)$,
to be generally nonlinear, we studied their
nontrivial fixed points. As mentioned earlier,
Lorenz function $L(x)$ has trivial fixed points
(at $x=0$ and $1$), while the Complementary
Lorenz function $L^{(c)}(x) \equiv 1-L(x)$ has
a nontrivial fixed point at $x = k$, the Kolkata
index~\cite{ghosh2014inequality}. It also offers a tangible
interpretation: $k$-index gives the fraction $k$
of the total wealth possessed by the rich
$(1 - k)$ fraction of the population and gives a
quantitative generalization of the Pareto's 80-20
law~\cite{wiki_pareto}. As discussed earlier, it can also
 be viewed as a normalized $h$-index of social
inequality.
Some unique features of Kolkata Index $(k)$ may be noted:
a) Gini and other indices are mostly some average
quantities  based on the Lorenz function $L(x)$, which
has trivial fixed points.  $k$ is a fixed point of the
Complementary Lorenz function $L^{(c)}(x)$ and if one
considers the simplest form of Lorenz function $L(x)
= x^2$, then $k = (\sqrt 5 - 1)/2$, inverse of the Golden
Ratio~\cite{chatterjee2017socio}. b) $k$ gives the fraction of wealth possessed by the rich
$1-k$ fraction of the population. As such, it provides a
quantitative generalization of the Pareto’s 80-20 law
(see e.g.,~\cite{wiki_pareto}). 
The observed values of $k$ index for most of the cases of social inequalities~\cite{ghosh2016inequality,chatterjee2017socio,sinha2019inequality,banerjee2020kolkata,banerjee2020social} seem to fall in the range 0.80-0.86 (though have much smaller values today for world economies, presumably because of various welfare measures).
c)  $k$-index is equivalent to a normalized version of the
Hirsch-Index ($h$).  While $h$ corresponds to the fixed point
of the  publication success rate (measured by the
integer numbers of citations) falling off nonlinearly
with number of papers by individual academicians, 
$k$ corresponds to the fixed point  (fraction) of $1 -  L(x)$,
where $L(x)$ gives the nonlinearly varying fraction of
cumulative wealth possessed by the increasing (from
poor to rich) fraction  of  population in any society. \\

\noindent
C) Kinetic Exchange Model of Income and Wealth Distributions

We have discussed already in section 2 some
details of the  Kinetic Exchange model of income and
wealth distributions. In an ideal gas, in thermal
equilibrium, the number density $n(\epsilon)$ of
(Newtonian) particles (atoms or molecules) having
kinetic energy  $\epsilon$ is given by

\begin{equation}
 n(\epsilon) = g(\epsilon)f(\epsilon, T) \sim \sqrt (\epsilon) \exp(-\epsilon/\Delta). \label{eq_secIII_I}
\end{equation}

\noindent
where $\Delta$ is a constant, the density of
states $g(\epsilon) \sim $ $\sqrt \epsilon$ (coming from
the counting of three dimensional momenta vectors
which correspond to the same kinetic energy) and
$f(\epsilon) \sim$ $\exp(-\epsilon/\Delta)$ (coming
from  stochastic, or entropy maximizing, scatterings
between the particles, conserving their total kinetic
energy). As discussed already in section 2, to
get the ideal gas equation of sate $PV = NKT$, where
$P$ and $V$ denote respectively the  pressure and
volume of the gas at absolute temperature $T$, by
calculating the pressure from the average transfer of
momentum of the particles per unit area of the container (using Eq.(~\ref{eq_secIII_I}),~\cite{saha1931treatise}),
one identifies, $\Delta = KT$.

Following similar arguments~\cite{saha1931treatise,dragulescu2000statistical} (see also~\cite{yakovenko2009colloquium}
), one gets  (as discussed in section
2)  for the steady state distribution of the number
$n(m)$ of agents in the market  with income or
wealth $m$.

\begin{equation}
  n(m) = g(m) f(m) = c \exp (-m/\Delta'). \label{eq_secIII_II}
\end{equation}

\noindent
where $f(m) \sim \exp(- m/\Delta')$, with $g(m)$ a constant $c$ (unlike in expression~(\ref{eq_secIII_I})),
and $\Delta'$ as constants for the trading market. This is
because, in a trading market, there is no production
(growth)  or decay and the total amount of money
(equivalent to energy in Kinetic theory of ideal gas) as
well as the number of traders (buyers and sellers) remain
fixed. Stochastic money exchanges in the trades involving
indistinguishable  buyers and sellers (who change their
roles in different trades),  keeping the buyer-seller
total money in any trade to remain constant, leads to a distribution given by expression~(\ref{eq_secIII_II}). This is  also because, there
can not be any equivalent of the particle momenta vectors
for the agents in the market and hence  the density
of states $g(m)$ here is a constant.  One must also have
$\int_0^M n(m)dm = N$, the total number of traders in the
market, and $\int_0^M mn(m)dm = M$, the total amount of
money. These give the effective temperature of the market $\Delta' = M/N$, the average money in circulation in the market (economy).

As documented in several books and reviews (see e.g.,~\cite{saha1931treatise,chatterjee2005econophysics,chatterjee2007kinetic}), 
the income or wealth
distributions in  any society have a Gamma function-like
dip near zero income or wealth (unlike in the exponential
distribution case discussed above, where the number density of
pauper is the maximum). Also, as is well known~\cite{chatterjee2007kinetic,chakrabarti2013econophysics} 
, the tail end of the distribution is known to
be much more fat, described by the Pareto power law, and
not by the thin exponentially decaying distribution.

As mentioned in the earlier section, following  Saha and
Srivastava's indication in their book~\cite{saha1931treatise}, we
explored how the kinetic theory of trading markets
indicated above could be extended to accommodate a
Gamma-like distribution  at the least and explore
further to capture the Pareto tail of such a
distribution as well.

We noted that many of the economics text books, in
their chapters on Trades, discuss about the saving
propensity of the traders (habit of saving a fraction
of the income or wealth possessed by the trader and
do trade with the rest). We immediately realized~\cite{chakraborti2000statistical,chatterjee2004pareto}, 
 if one introduces the saving propensity
of each trader, so that each trader saves a fraction
of their individual money (wealth) before the trade
and allows (random) exchanges of the rest amount in the
trade (keeping  the total money or wealth, including
the saved portions, conserved), the traders will never
become pauper. Unlike in the random exchange case (as
in kinetic theory of gases, where one trader may lose
its entire money or wealth to the other in any trade),
here to lose the entire money acquired at any point of time, the trader has to lose
every time after that as the trader continues the successive trades (and consequently the saved portion
becomes infinitesimal). The number density of paupers
(having zero wealth) will become zero for any
non-vanishing saving fraction of the traders and
the exponential distribution will become unstable
and the resulting steady  state distributions will
capture the above mentioned  desirable features. This
is a non-perturbative result; any amount of saving
by the traders will induce this feature.

With uniform  saving the exponential  distribution
collapses and the stable distribution becomes
Gamma-like~\cite{chakraborti2000statistical} (see also~\cite{chakrabarti2009microeconomics} for a
micro-economic derivation of the kinetic exchange
equations from the Cobb-Dauglas utility maximization
with money saving propensity of the traders, and~\cite{quevedo2020non}
 for extended microeconomic formulation of
Kinetic exchange models, having economic growths,
by incorporating additional saving of the production
in the utility maximization equation). The steady state
distribution becomes initially Gamma-like but crossing
over to Pareto-like power-law decay when traders have
non-uniform saving propensities~\cite{chatterjee2004pareto}. The saving
propensity magnitudes determine the most-probable
income (wealth) and the income (wealth) crossover
point for Pareto tail of the distribution (see~\cite{pareschi2013interacting,ribeiro2020income,chakrabarti2013econophysics} for details).

It may be mentioned in this connection, that
one kind of saving by the traders, considered
early by our group (including the students Anirban
Chakrabarti and Srutarshi Pradhan),  can in fact
lead to wealth condensation or extreme inequality.
When two randomly selected traders agree to trade
(in the so called 'Yard Sale' trade mode), such that the
richer one among them will retain or save the extra
money or wealth compared to that of her trade
partner, the dynamics will eventually lead
to aggregation of the entire money or wealth in
the hand of one trader and the dynamics will stop. This happens because once any trader
becomes pauper (looses entire money or wealth), no
other trader (with money) will engage in trade with
her. Although this Yard Sale model has this
uninteresting wealth condensation feature, it showed
some interesting slow dynamics and Anirban published
that result~\cite{chakraborti2002distributions}. Later, it was shown that inclusion
of tax in the  model, in the sense that a fraction of
money is collected by the Government (non-playing
member of the system) in  every trade and after some
period of collections, redistributes the money among
all (by investing on general social facilities, like
road, hospital, etc constructions, used equally by all
in the society). Because of this general upliftment,
the paupers come back to the trades and  interesting
steady state money distribution can emerge and such
models of wealth distribution have become an
active area of research (see e.g.,~\cite{boghosian2019} for a
popular review on this development).

Kinetic model of gases and the kinetic theory is the
first and extremely successful many-body theory in
physics. Economic systems, markets in particular, 
are intrinsically many-body dynamical systems.
Kinetic exchange models of markets may therefore
be expected to provide the most successful models of market systems. In the kinetic exchange model, when one of the trader of a randomly chosen pair of traders is deliberately the poorest one at that instant of time (trade), the dynamics induces a self-organization in the market such that a `poverty line'   is spontaneously developed so that none of the trader remains below the emerged (self-organized) poverty threshold (see~\cite{iglesias2010simple,ghosh2011threshold} and references therein). \\

\noindent
D) Statistics of the Kolkata Paise Restaurant Problems

Kolkata had, long back, very cheap fixed price `Paise
Restaurants' (also called `Paise Hotels'; Paise is,
rather was, the smallest  Indian coin). These `Paise 
Restaurant' were very popular among the daily laborers
in the city. During lunch hours, these laborers used to
walk down (to save the transport costs) from their place
of work to one of these  restaurants. These 
restaurants would prepare every day a (small) number of
such dishes, sold at a fixed price (Paise). If several
groups of laborers would arrive any day to the same
restaurant, only one group would get their lunch
and others would miss their lunch that day. There were
no cheap communication means in those days (like mobile
phones) for mutual communications, for deciding 
the respective restaurants. Walking down to the next
restaurant would mean failing to report back to work on
time! To complicate this collective learning and decision
making problem, there were indeed some well-known rankings
of these restaurants, as some of them and would offer
tastier items compared to the others (at the same
cost, paise, of course) and people would prefer to
choose the higher rank of the restaurant, if not crowded!
This `mismatch' of the choice and the consequent decision
not only creates inconvenience for the prospective
customers (going without lunch), would also mean `social
wastage' (excess unconsumed food, services or supplies
somewhere).

A similar problem arises when the public administration
plans and provides hospitals (beds) in different localities,
but the local patients prefer `better' perceived hospitals
elsewhere. These `outsider' patients would then have to
choose other suitable hospitals elsewhere. Unavailability
of the hospital beds in the over-crowded hospitals may
be considered as insufficient service provided by the
administration, and consequently the unattended potential
services will be considered as social wastage.

This kind of games~\cite{chakrabarti2009kolkata} (see~\cite{chakraborti2015statistical,chakrabarti2017econophysics} for
recent reviews), anticipating the possible strategies of
the other players and acting accordingly, is very common
in society. Here, the number of choices need  not be very
limited (as in standard binary-choice formulations of most
of the games, for example in Minority Games~\cite{chakrabarti2017econophysics,challet2013minority,chakraborti2015statistical} and
the number of players can be truly large! Also, these are
not necessarily one shot games, rather the players can
learn from past mistakes and improve on their selection
strategies for choosing the next move. These features make
the games extremely intriguing and also versatile, with major
collective or emerging social structures, not comparable to
the standard finite choice, non-iterative games among finite
number of players. Such repetitive collective social
learning for a community sharing past knowledge for the
individual intention to be in minority choice side in
successive attempts are modeled by the `Kolkata Paise
Restaurant' (KPR) problem or, in short, by the `Kolkata
Restaurant' problem.

KPR is a repeated game, played among a large
number  of players or agents having no simultaneous
communication or interaction  among themselves. In KPR,
the prospective players (customers/agents) choose from
restaurants each day (time) in parallel decision
mode, based on the past (crowd) information and their
own (evolved or learned) strategies. There is no budget
constraint to restrict the choice (and hence the
solutions). Each restaurant has the same price for a
meal but having a different rank, agreed upon by
all the customers or players.

For simplicity, we may assume that each restaurant
can serve only one customer (generalization to
any fixed number of daily services for each would not
change the complexion of the problem or game). If
more than one customer arrives at any restaurant on
any day, one of them is randomly chosen and is served
and the rest do not get meal that day. Information
regarding the prospective customer or crowd distributions
for the earlier days (up to a finite memory size) is
made available to everyone. Each day, based on own
learning and the developed (often mixed) strategies, each
customer chooses a restaurant independent of the others.
Each customer wants to go to the restaurant with the
highest possible rank while avoiding a crowd so as to be
able to get the meal there. Both from individual success and
social efficiency perspective, the goal is to `learn
collectively' to utilize effectively the available
resources.

\begin{figure*}[htb]
\begin{center}
\includegraphics[width=6.5cm]{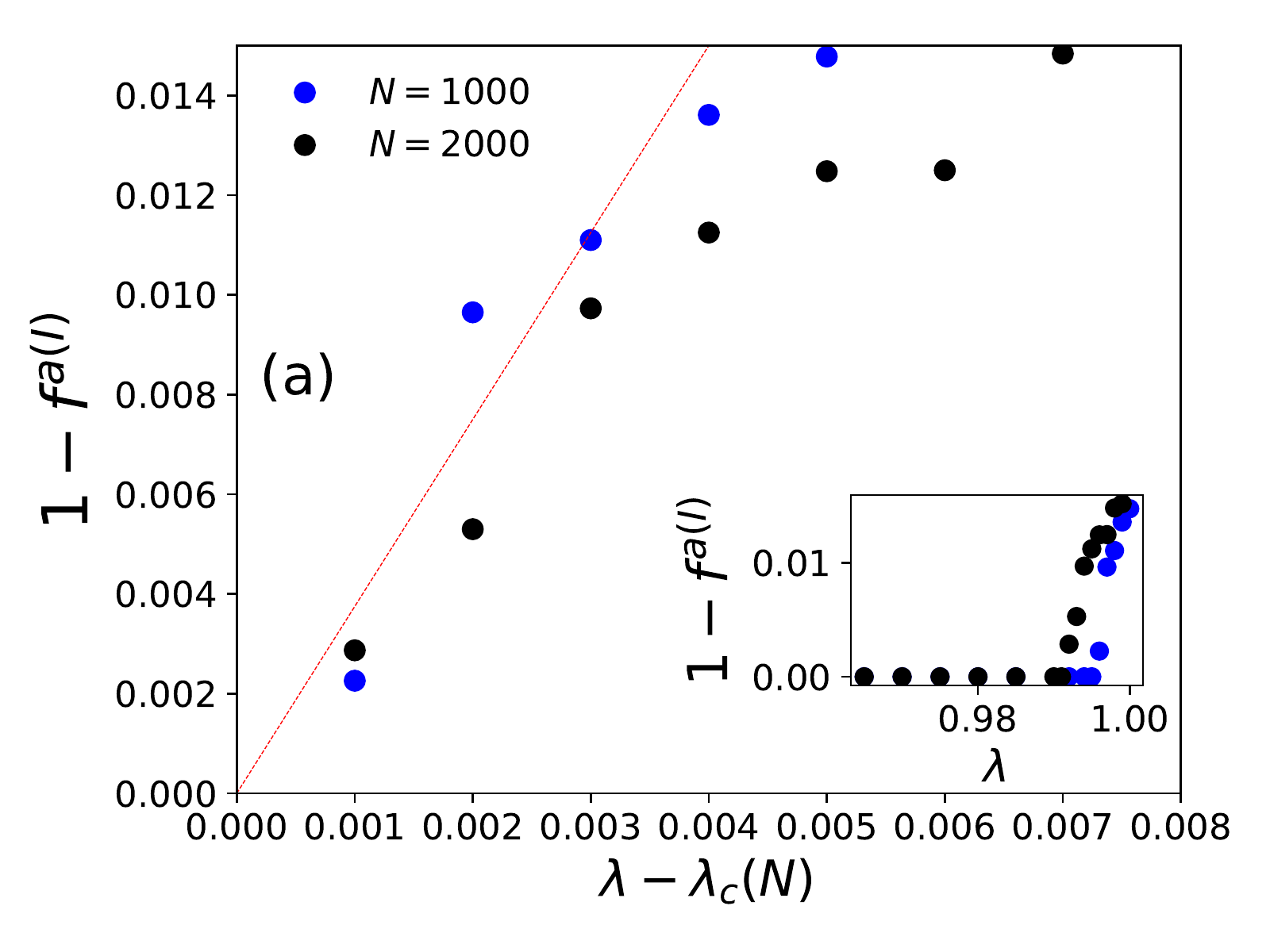}
\includegraphics[width=6.5cm]{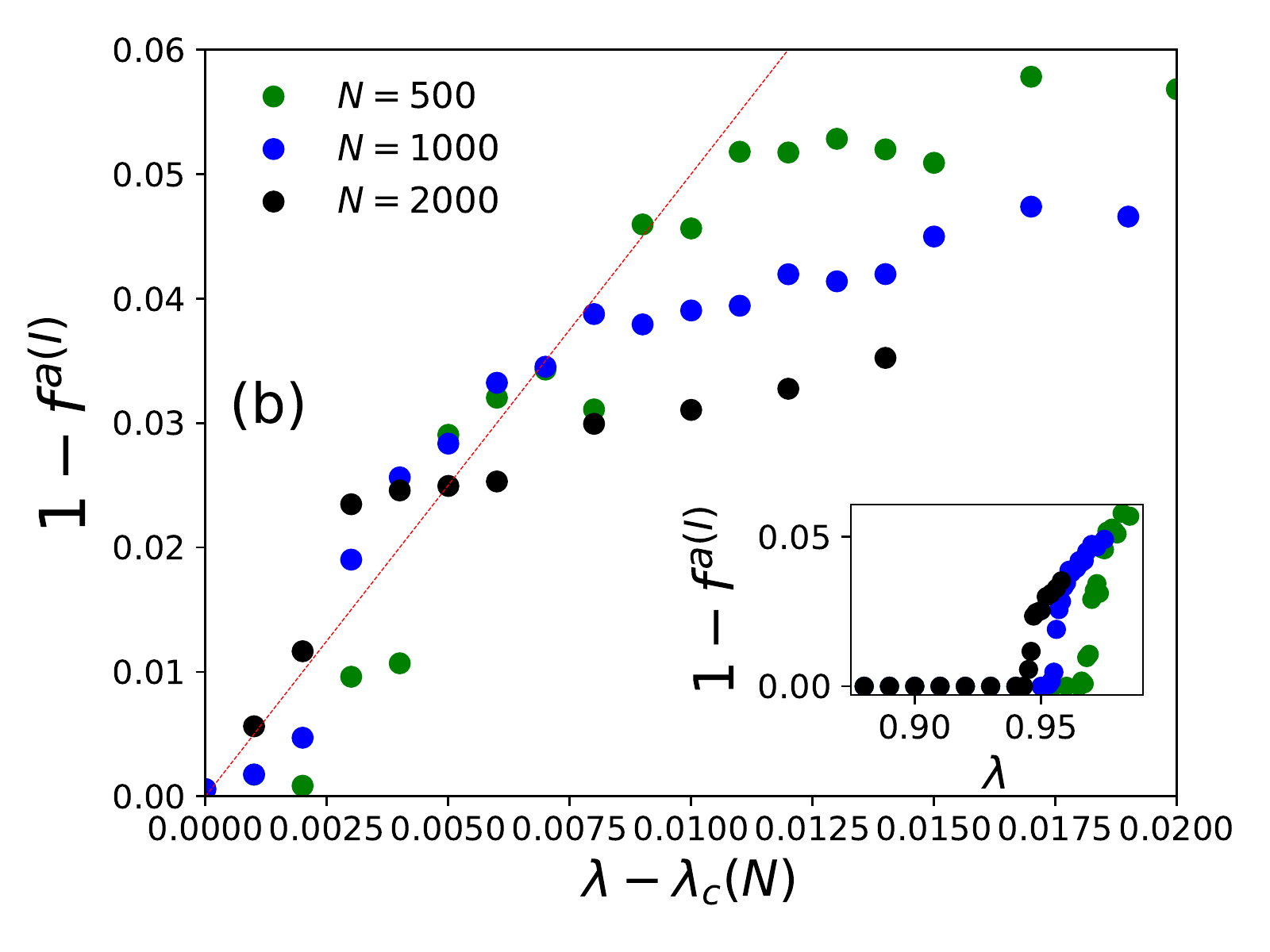}
\vskip 0.1cm
\includegraphics[width=6.5cm]{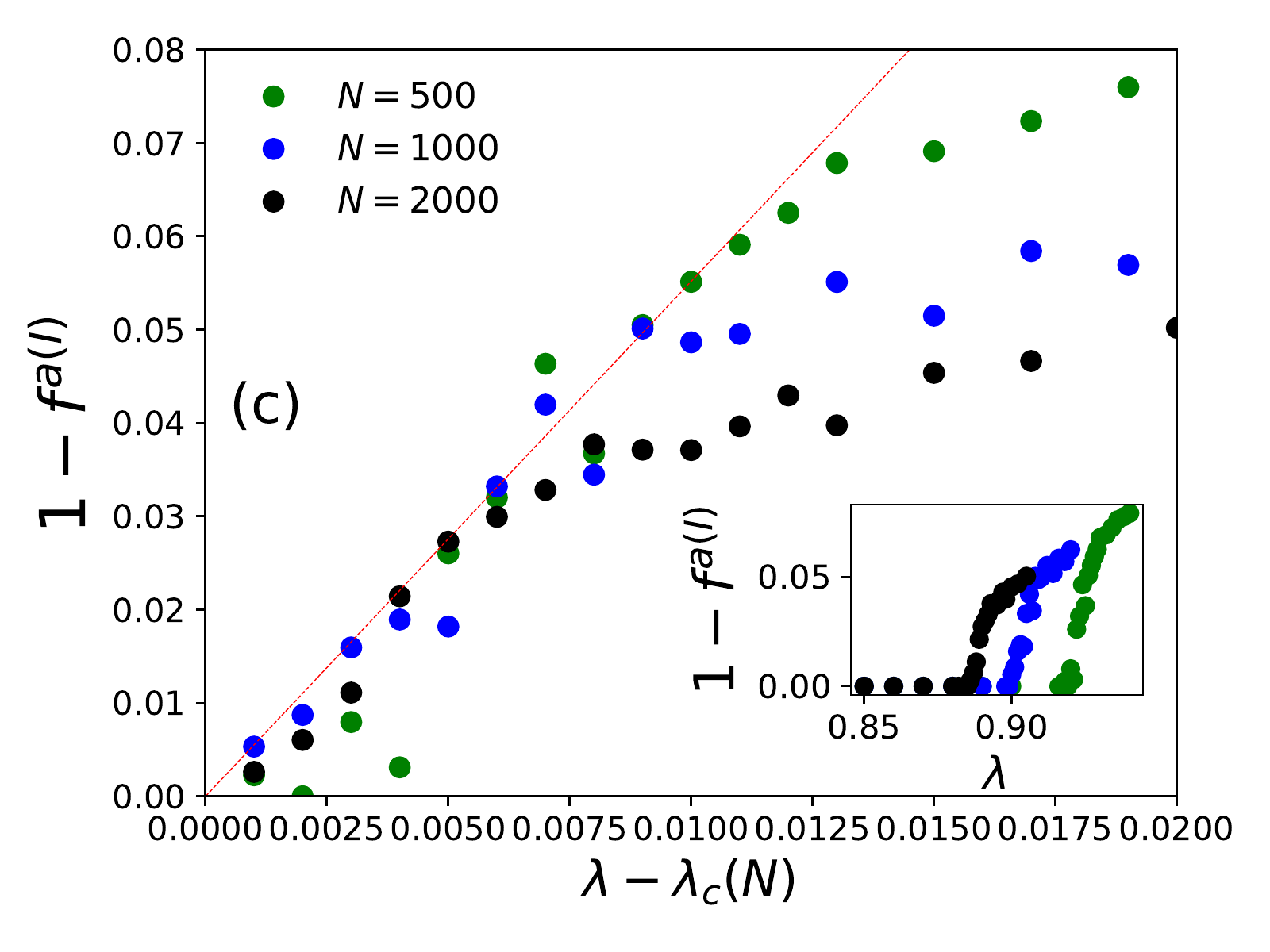}
\includegraphics[width=6.5cm]{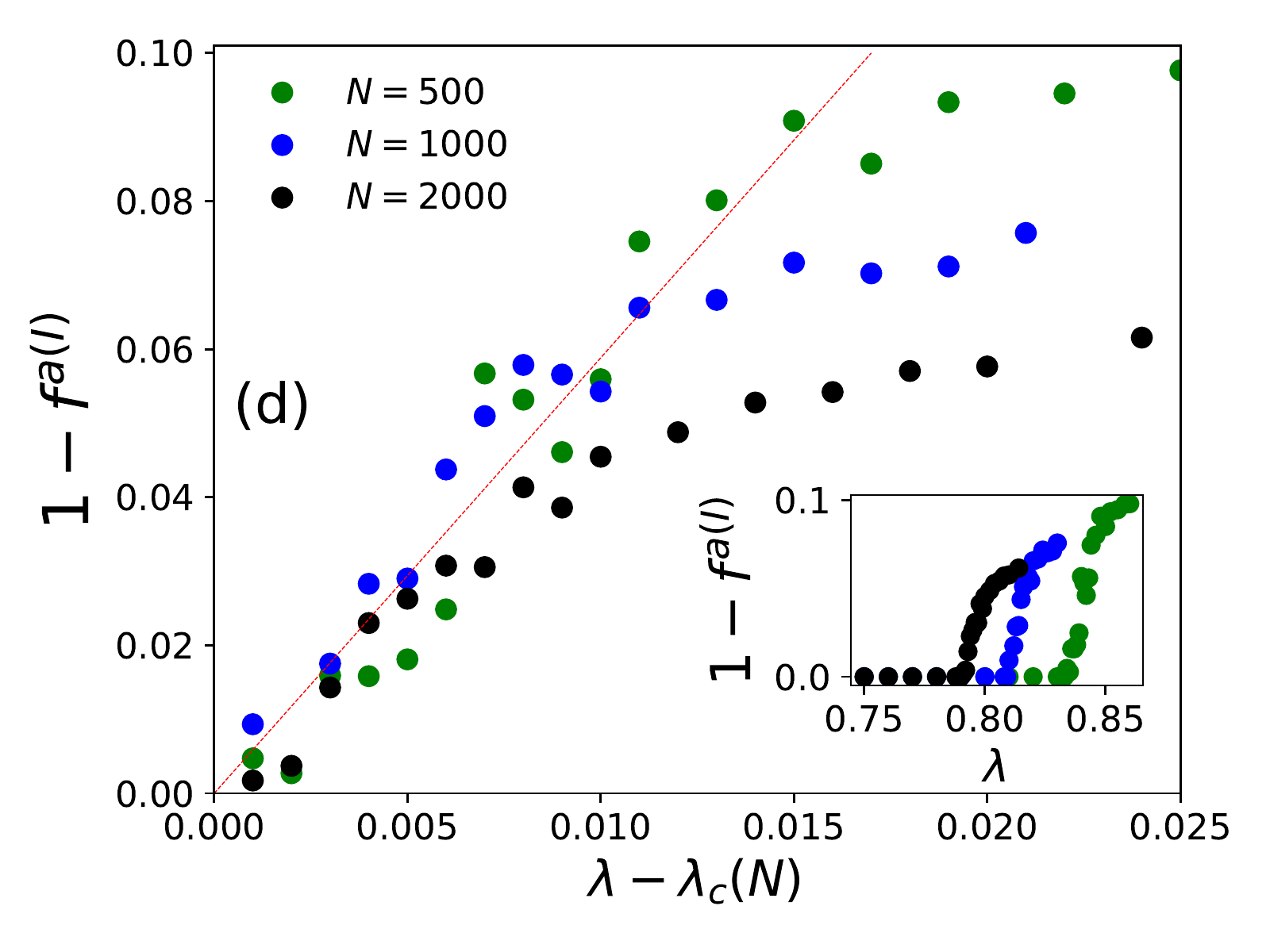}
\end{center}
\caption{ Plots of $(1-f^{a(I)})$ against $\lambda - \lambda_c(N)$ following strategy I at (a) $\alpha = 0.05$, (b) $\alpha = 0.25$, (c) $\alpha = 0.5$, (d) $\alpha = 1.0$. A power law holds for $(1-f^{a(I)})\sim(\lambda - \lambda_c(N))^{\beta}$ where $\beta = 1.0\pm0.05$. The insets show direct relationship between $(1-f^{a(I)})$ and $\lambda$ (for strategy I). }
\label{fig_secIV_I}
\end{figure*}

\begin{figure*}[htb]
\begin{center}
\includegraphics[width=6.5cm]{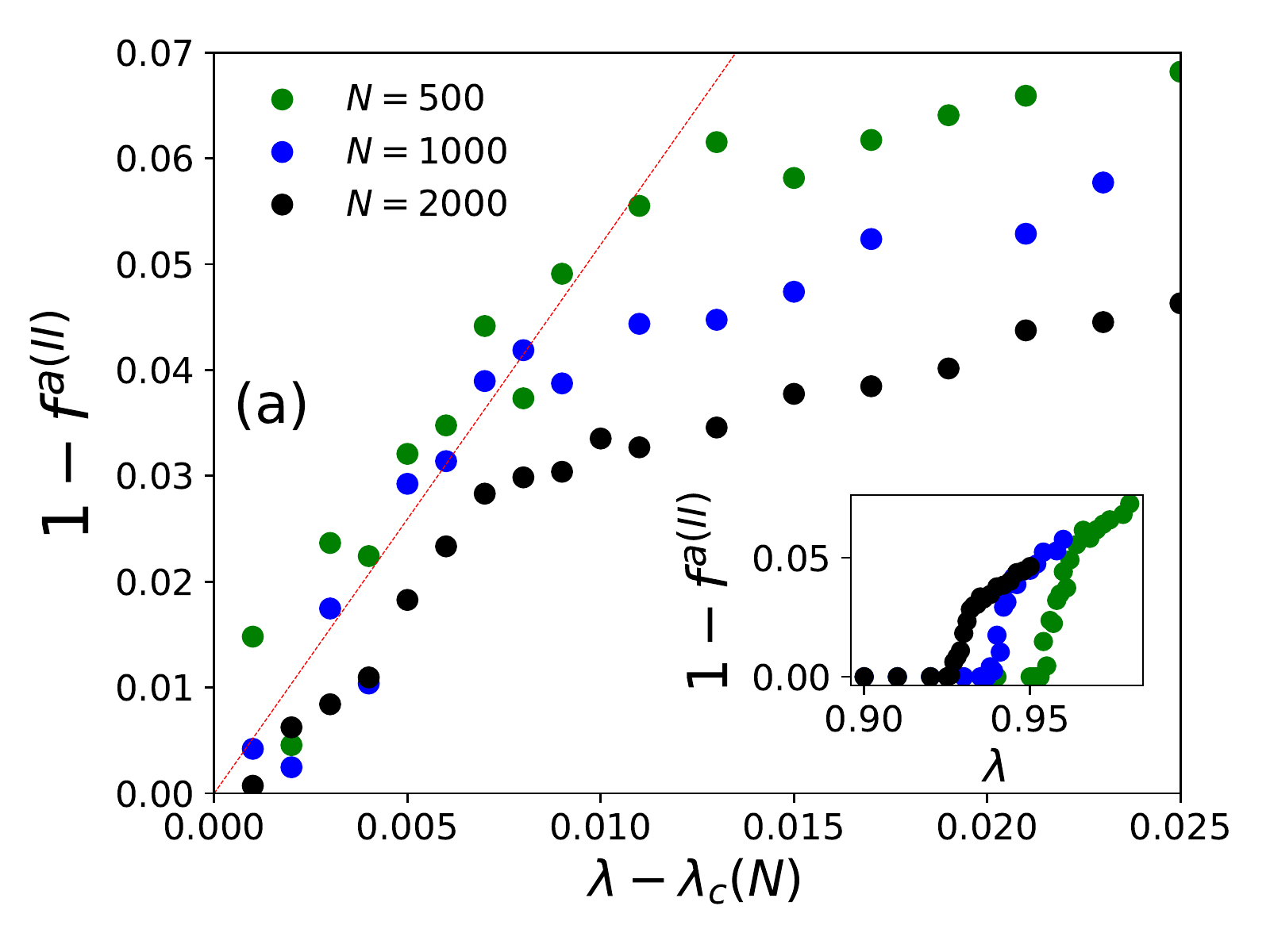}
\includegraphics[width=6.5cm]{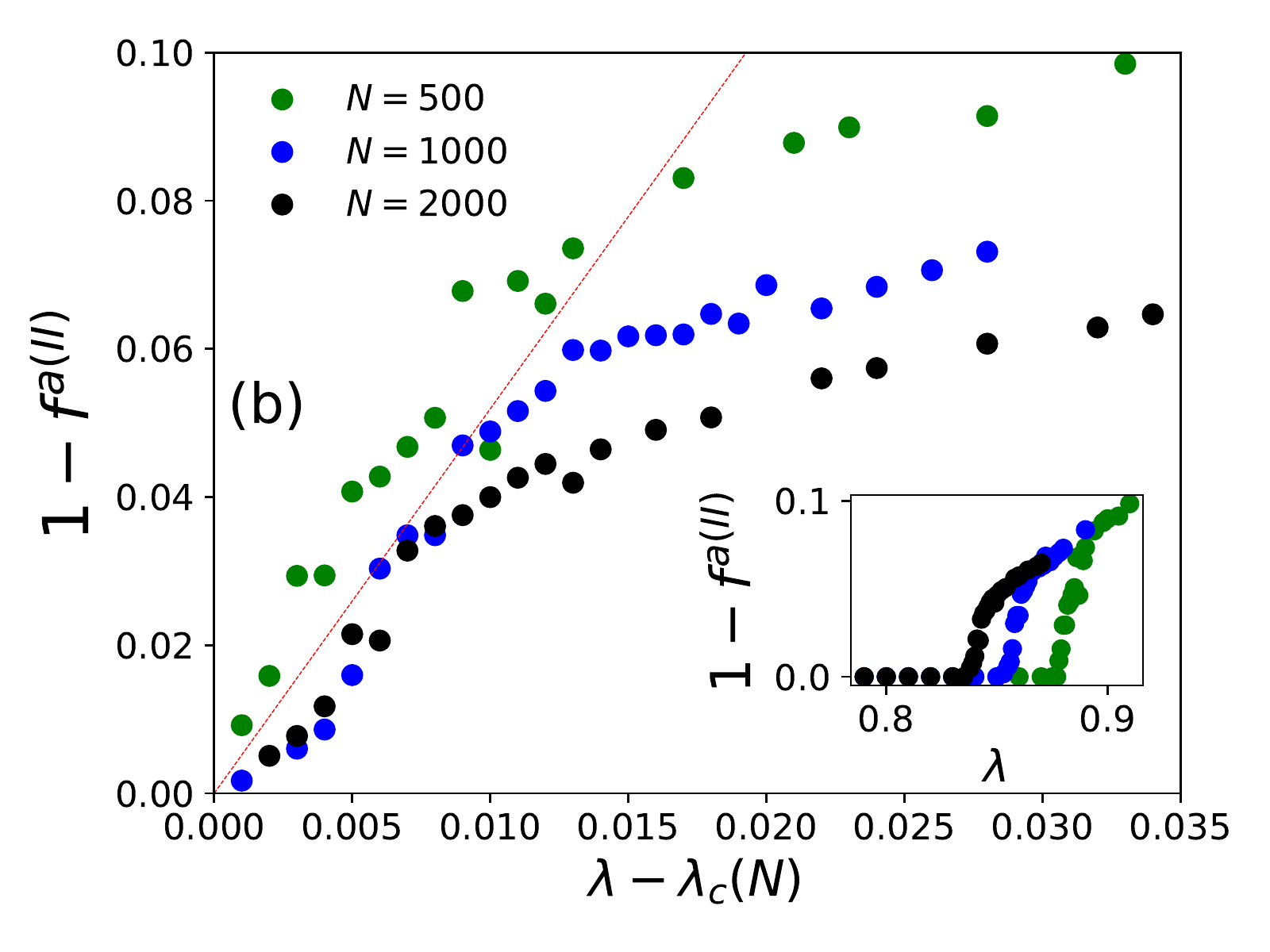}
\vskip 0.1cm
\includegraphics[width=6.5cm]{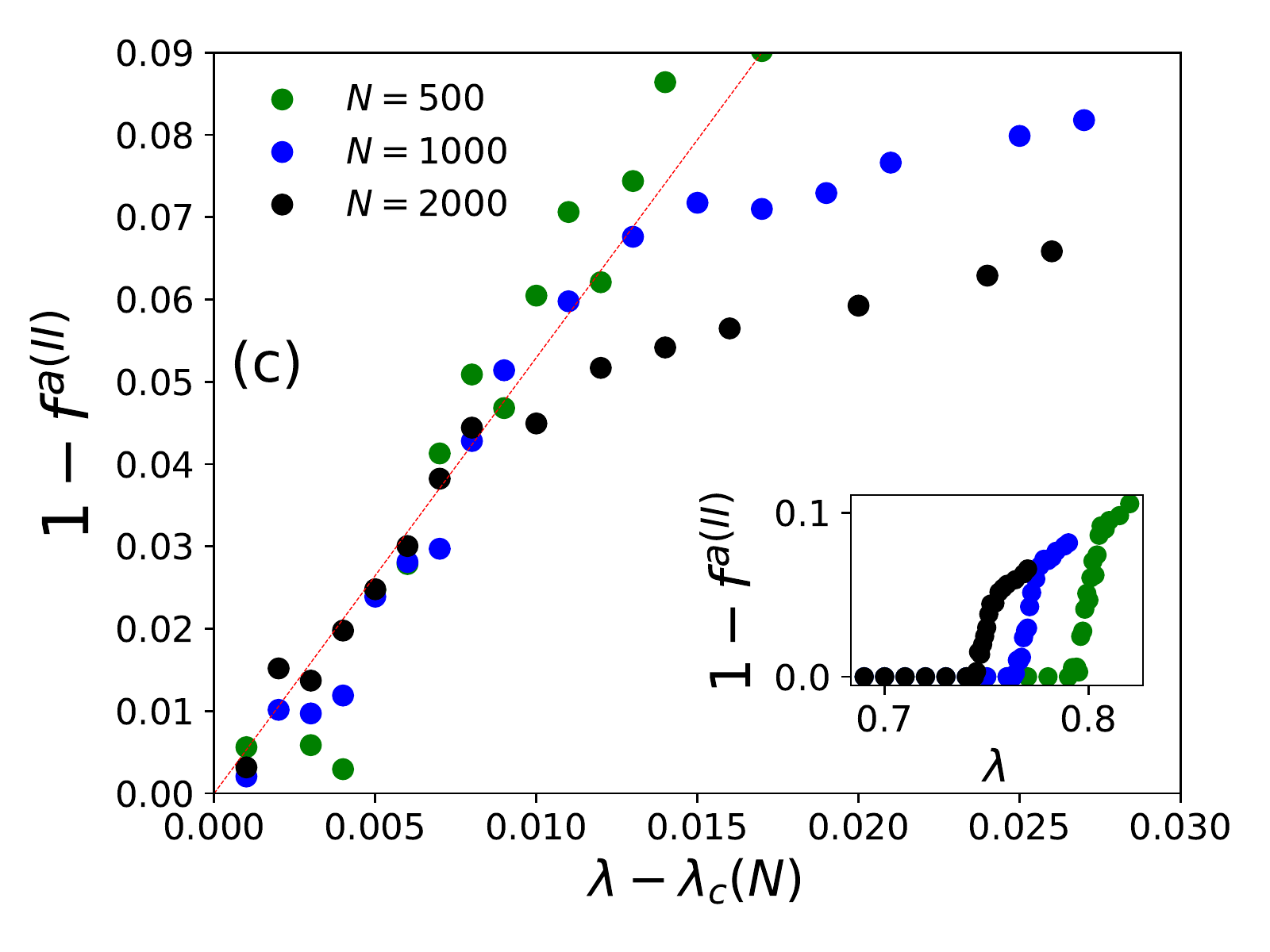}
\includegraphics[width=6.5cm]{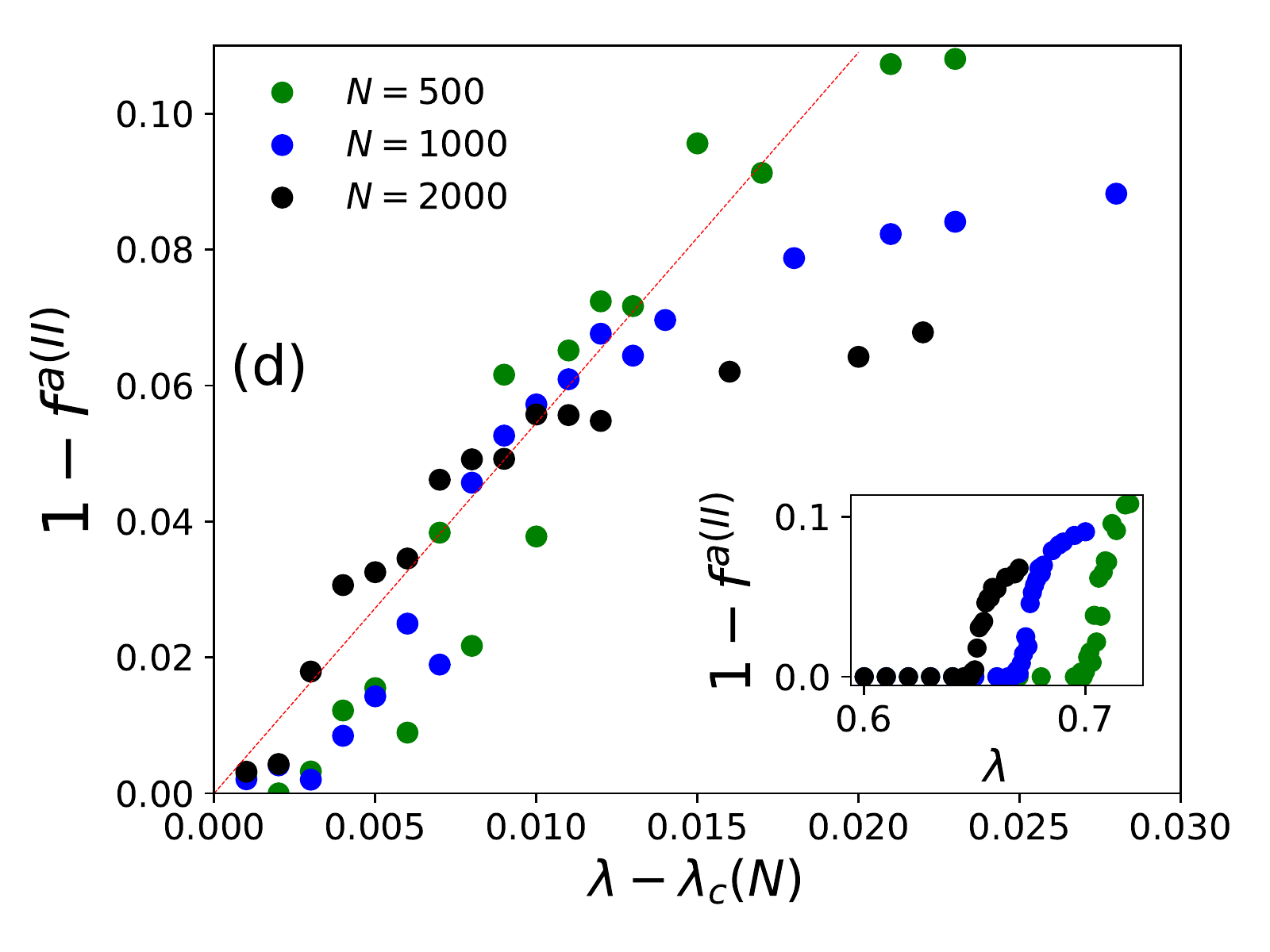}
\end{center}
\caption{ Plots of $(1-f^{a(II)})$ versus $\lambda - \lambda_c(N)$ following strategy II at (a) $p = 0.8$, (b) $p = 0.6$, (c) $p = 0.4$, (d) $p = 0.2$. A power law holds for $(1-f^{a(II)})\sim(\lambda - \lambda_c(N))^{\beta}$ with $\beta = 1.0\pm0.05$. The insets show direct relationship between variations of $(1-f^{a(II)})$ against $\lambda$ (for strategy II).
}
\label{fig_secIV_II}
\end{figure*}

The KPR problem seems
to have a trivial solution: suppose that somebody, say a
dictator (who is not a player), assigns a restaurant to
each person the first day and asks them to shift to the next restaurant
cyclically, on successive days. The fairest and most
efficient solution: each customer gets food on each day
(if the number of plates or choices is the same as that of
the customers or players) with the same share of the
rankings as others, and that too from the first day
(minimum evolution time). This, however, is NOT a true
solution of the KPR problem, where each customer or agent
decides on his or her own every day, based on
complete information about past events. In KPR, the
customers try to evolve learning strategies to eventually
to arrive at the best possible solution (close to the
dictated solution indicated above). The time for this
evolution needs also to be optimized;  for example, a
very efficient strategy, having convergence time which
grows with the number of players (even linearly), is
unsuitable for most of the social games, as our
life-span is finite, and (in a democracy) the number
of players or competitors can not be restricted or
bounded.

There have been many limiting formulations and studies
using tricks from statistical physics and quantum
physics (see e.g.,~\cite{chakraborti2015statistical,chakrabarti2017econophysics,ghosh2010statistics,
sharif2011quantum,sharif2013introduction,ghosh2017emergence,banerjee2018econophysics,sharma2018econophysics,
tamir2018econophysics,sinha2020phase}) and generalizations  in
computer science (see e.g.,~\cite{park2017kolkata}) and mobility
(vehicle on hire) markets (see e.g.,~\cite{martin2019,martin2017vehicle}
). We will present briefly in the next
section some specific  results of a new study
on the nature phase transition and resource utilization
in KPR with number of customers less (still very large)
than the number of restaurants.

\section{Some New Results for Statistics of the KPR Problem}

\begin{figure*}[htb]
\begin{center}
\includegraphics[width=6.5cm]{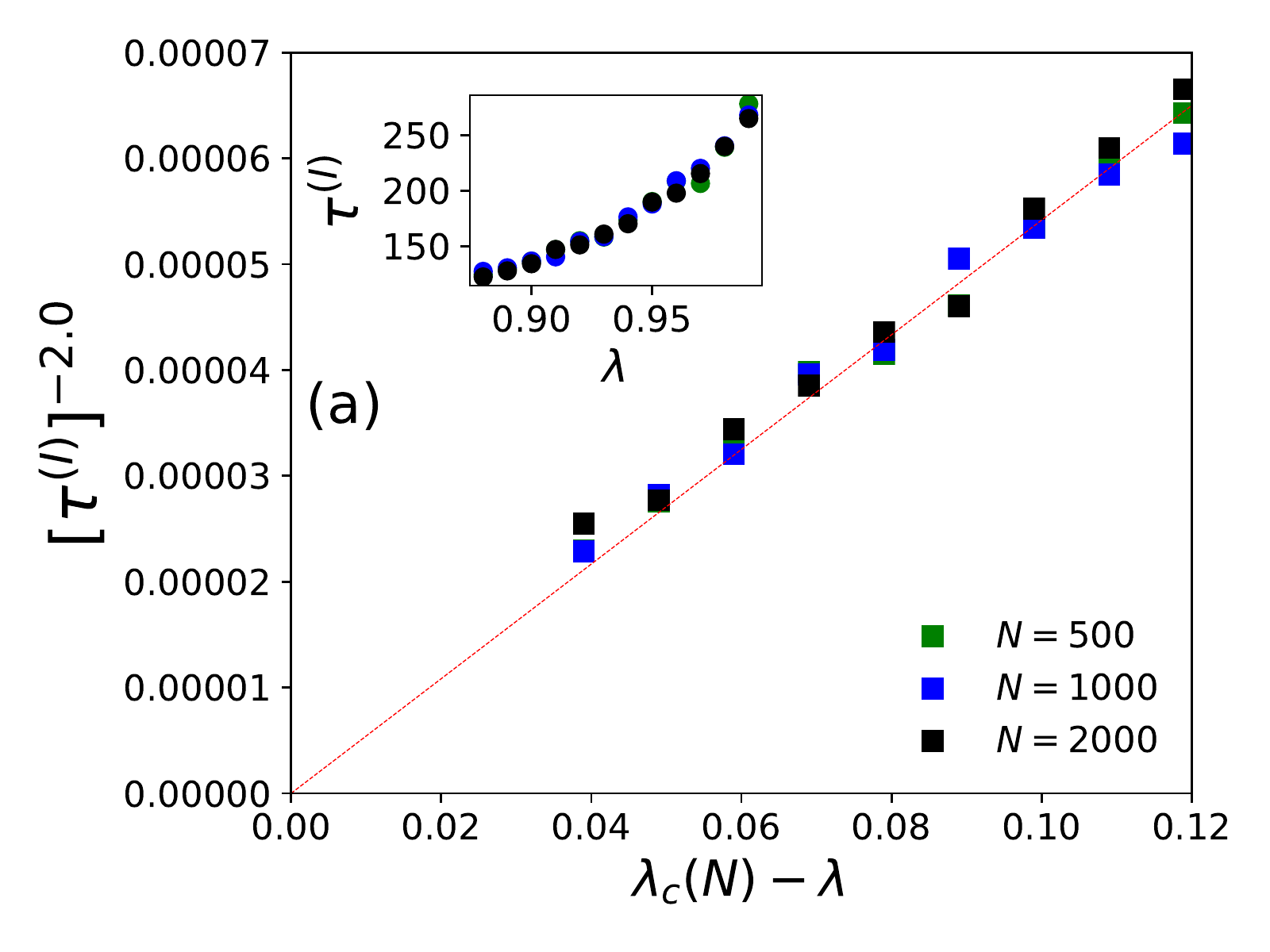}
\includegraphics[width=6.5cm]{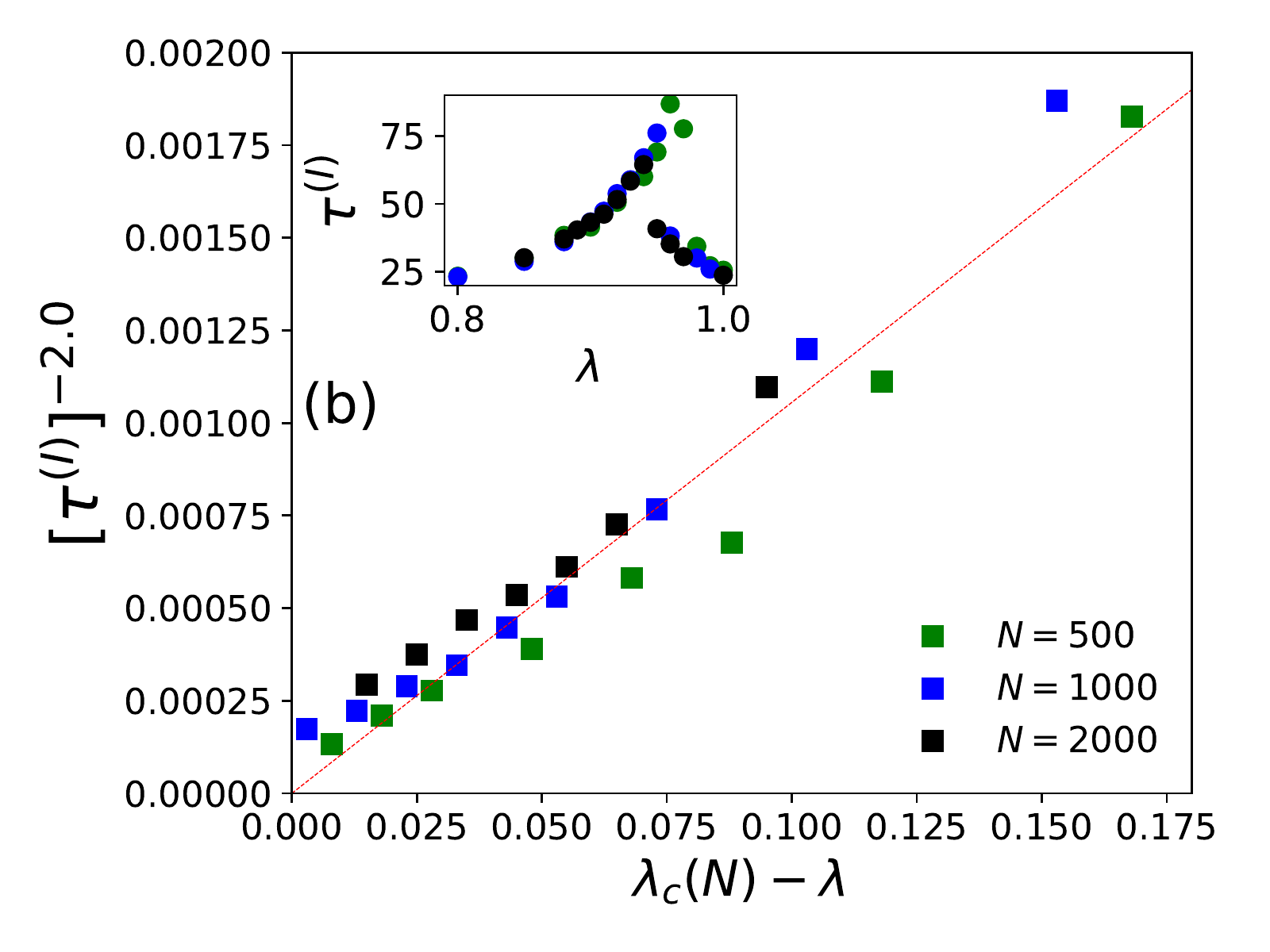}
\vskip 0.1cm
\includegraphics[width=6.5cm]{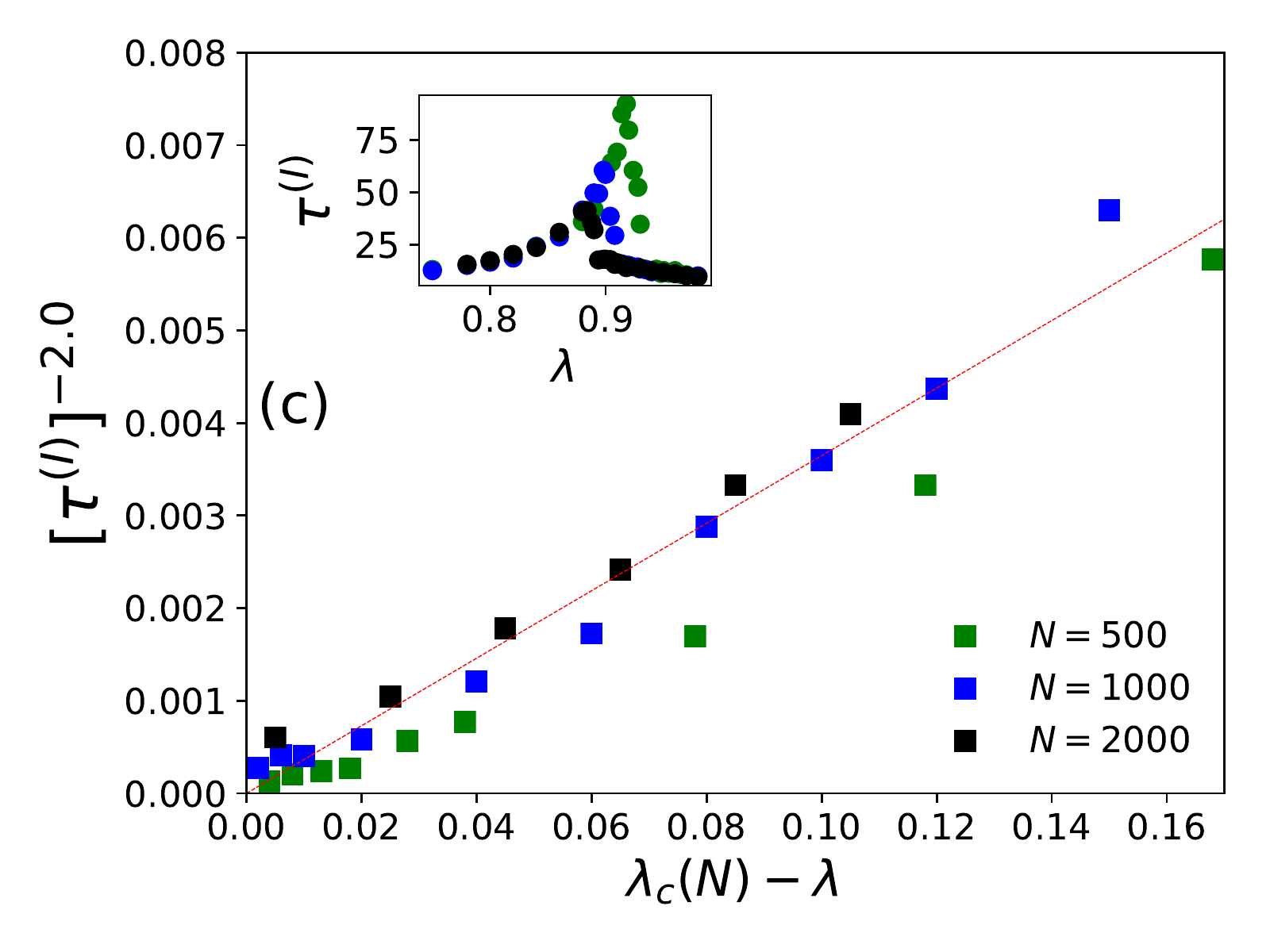}
\includegraphics[width=6.5cm]{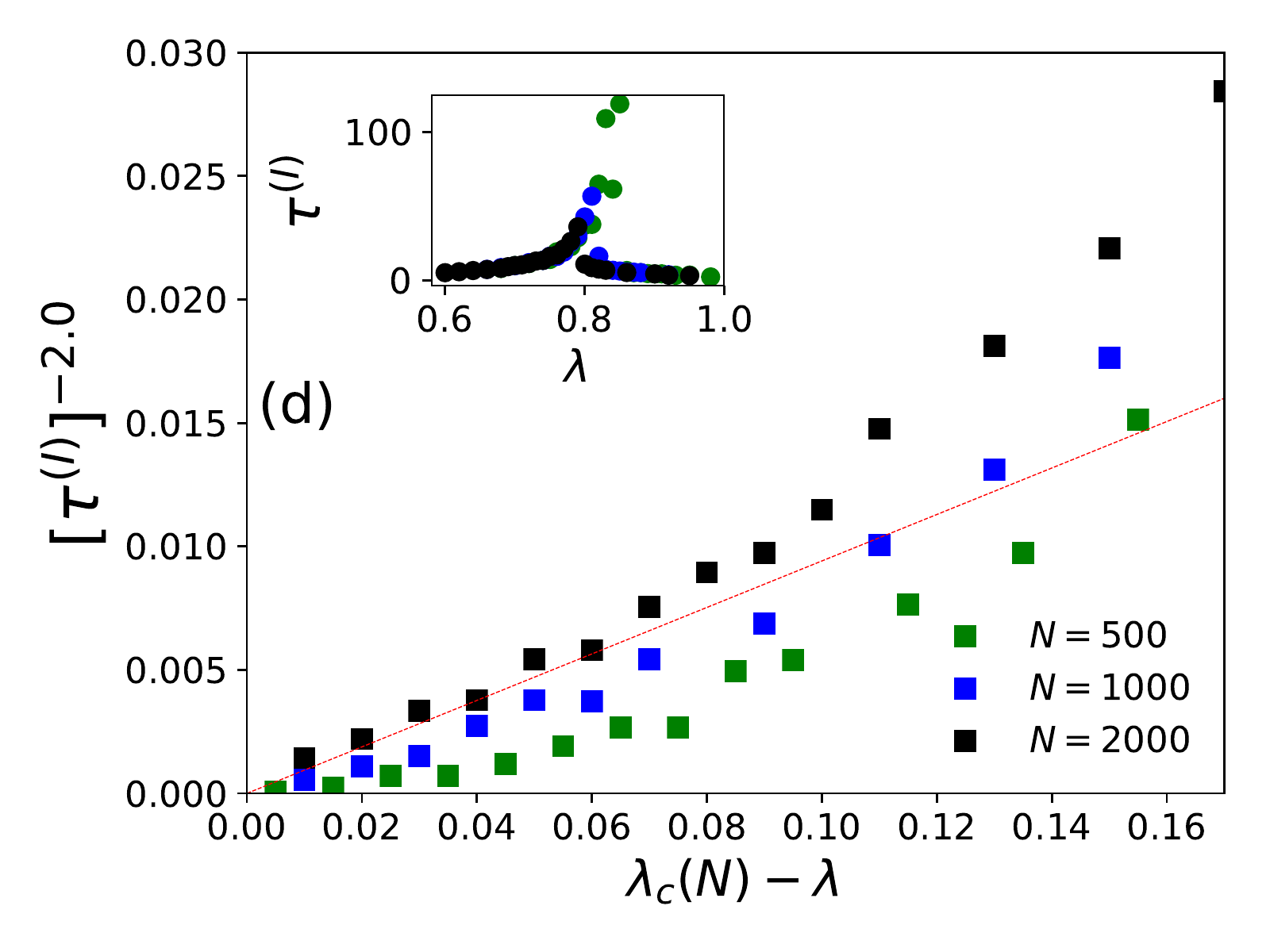}
\end{center}
\caption{ Plots of steady state convergence time $\tau^{(I)}$ from strategy I against $\lambda_c(N) - \lambda$ at (a) $\alpha = 0.05$, (b) $\alpha = 0.25$, (c) $\alpha = 0.5$, (d) $\alpha = 1.0$. A power law holds for $\tau^{(I)}\sim(\lambda_c(N) - \lambda)^{-\gamma}$ where $\gamma = 0.5\pm0.05$. The insets plot direct relationship between $\tau^{(I)}$ and $\lambda$ for different system sizes (for strategy I), also showing the variation of $\lambda$ as $\alpha$ increases. }
\label{fig_secIV_III}
\end{figure*}

\begin{figure*}[htb]
\begin{center}
\includegraphics[width=6.5cm]{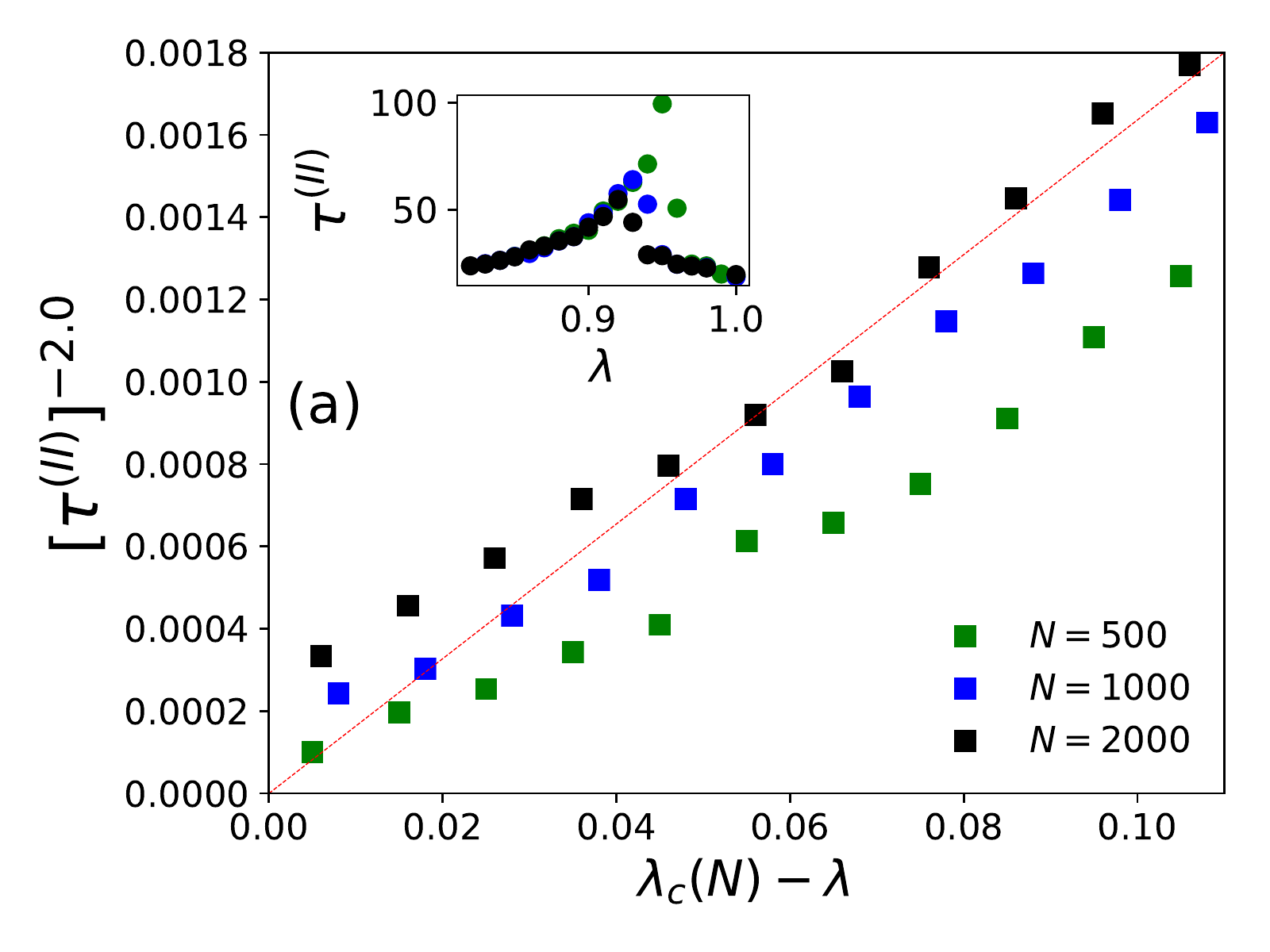}
\includegraphics[width=6.5cm]{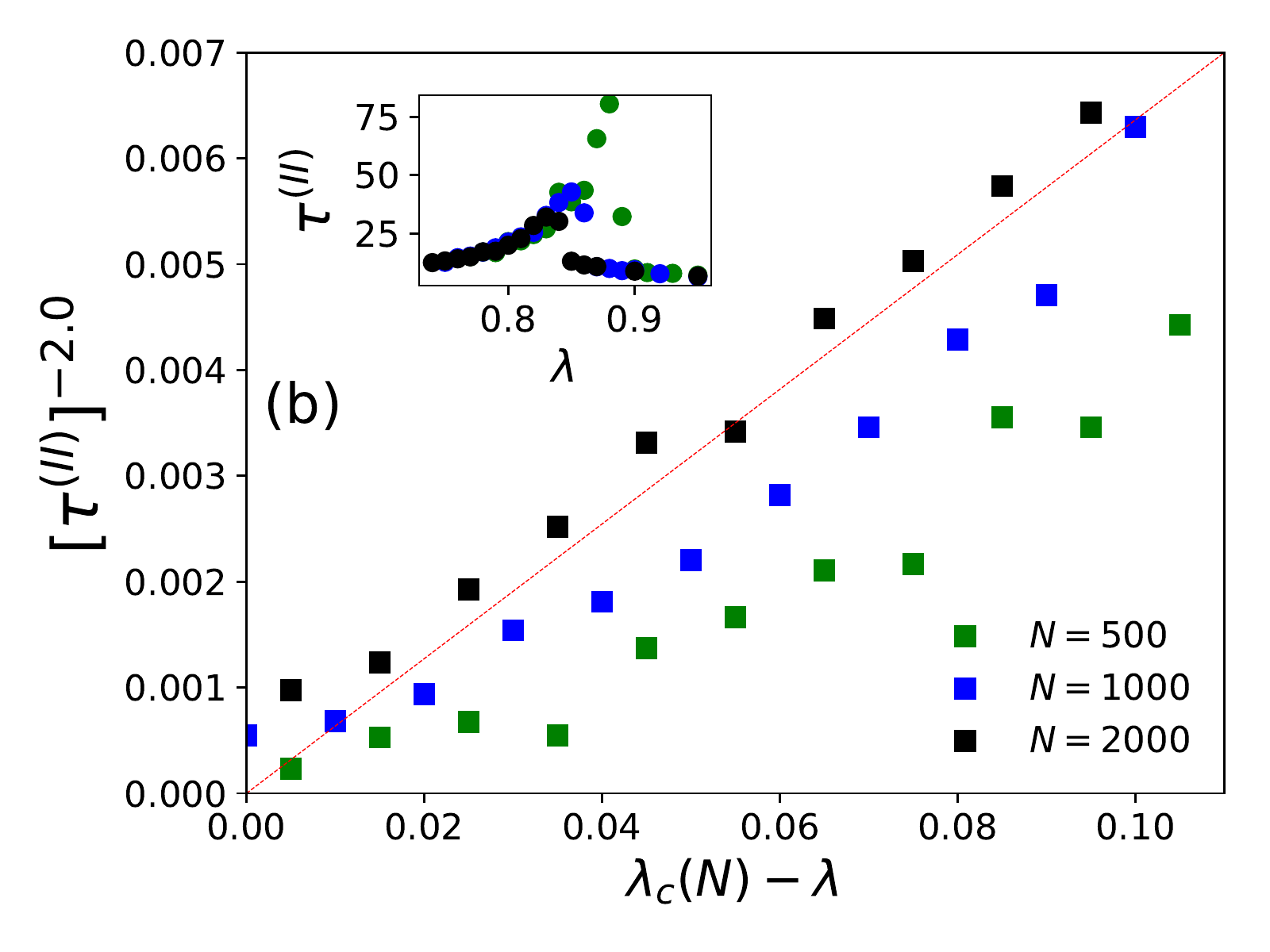}
\vskip 0.1cm
\includegraphics[width=6.5cm]{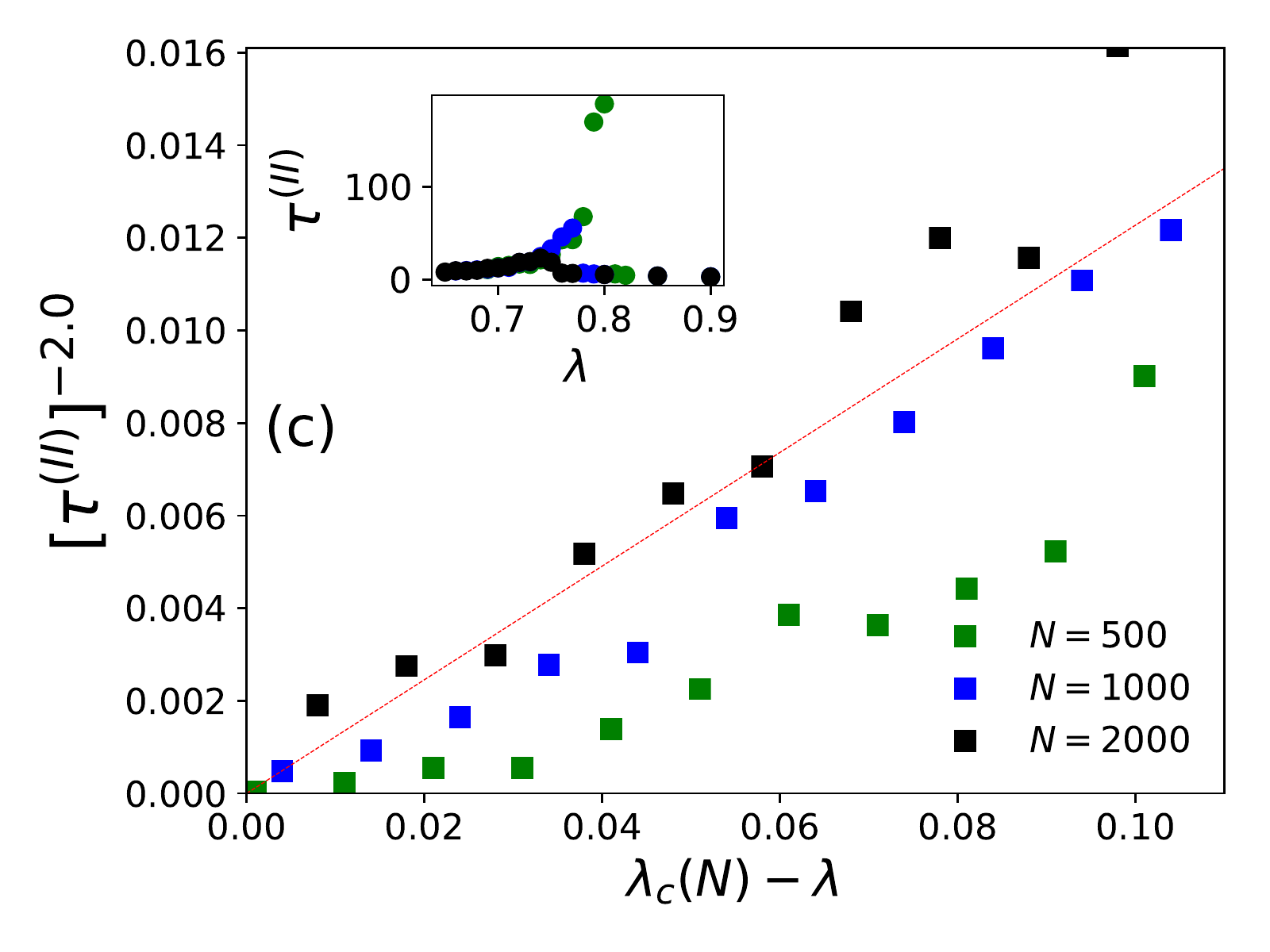}
\includegraphics[width=6.5cm]{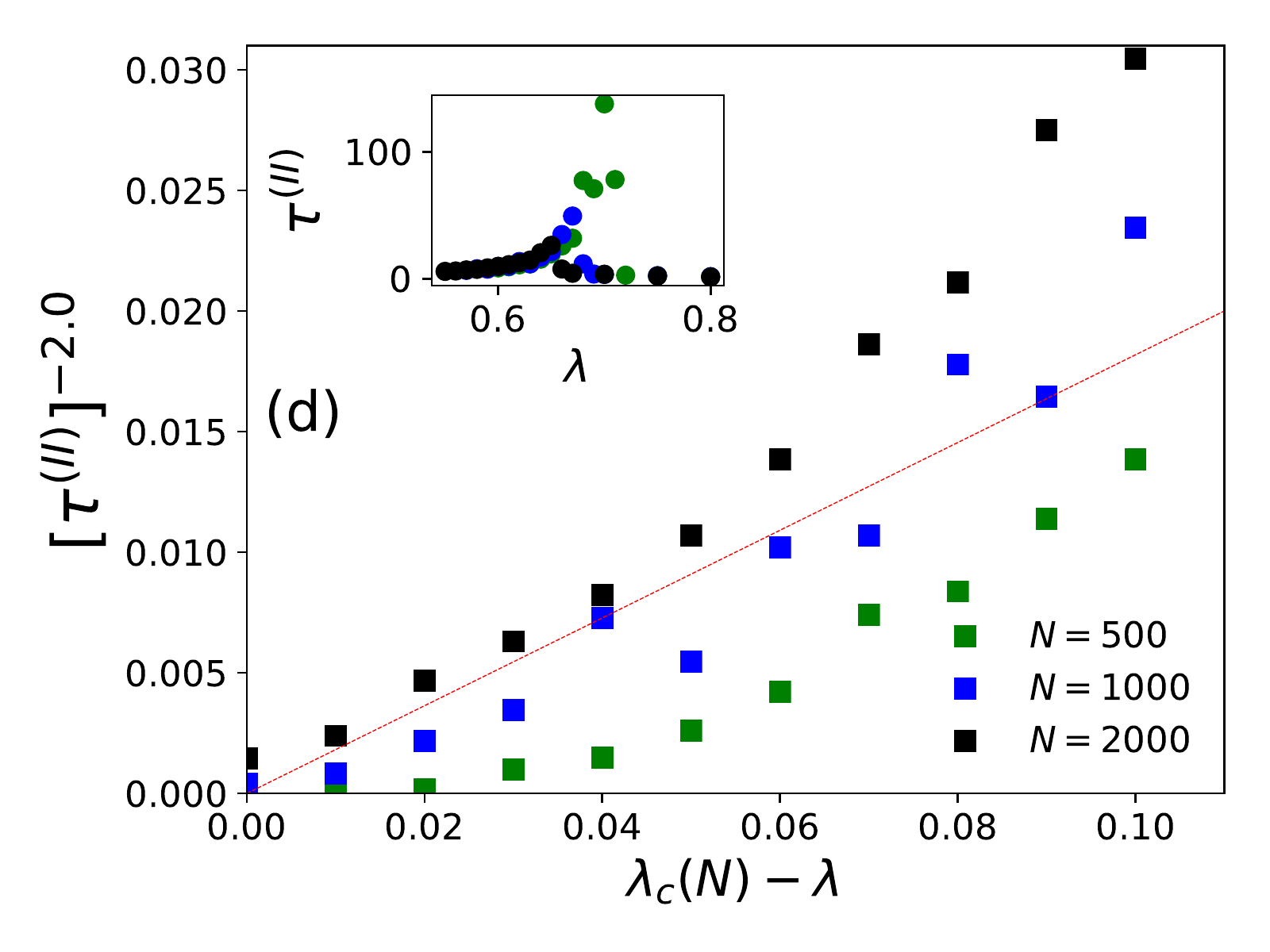}
\end{center}
\caption{ Plots of steady state convergence time $\tau^{(II)}$ against $\lambda_c(N) - \lambda$ following strategy II at (a) $p = 0.8$, (b) $p = 0.6$, (c) $p = 0.4$, (d) $p = 0.2$. A power law holds for $\tau^{(II)}\sim(\lambda_c(N) - \lambda)^{-\gamma}$ with $\gamma = 0.5\pm0.07$. The insets give direct relationship between $\tau^{(II)}$ and $\lambda$ for different system sizes (for strategy II), also showing the variation of $\lambda$ as $p$ decreases. }
\label{fig_secIV_IV}
\end{figure*}

\begin{figure*}[htb]
\begin{center}
\includegraphics[width=6.5cm]{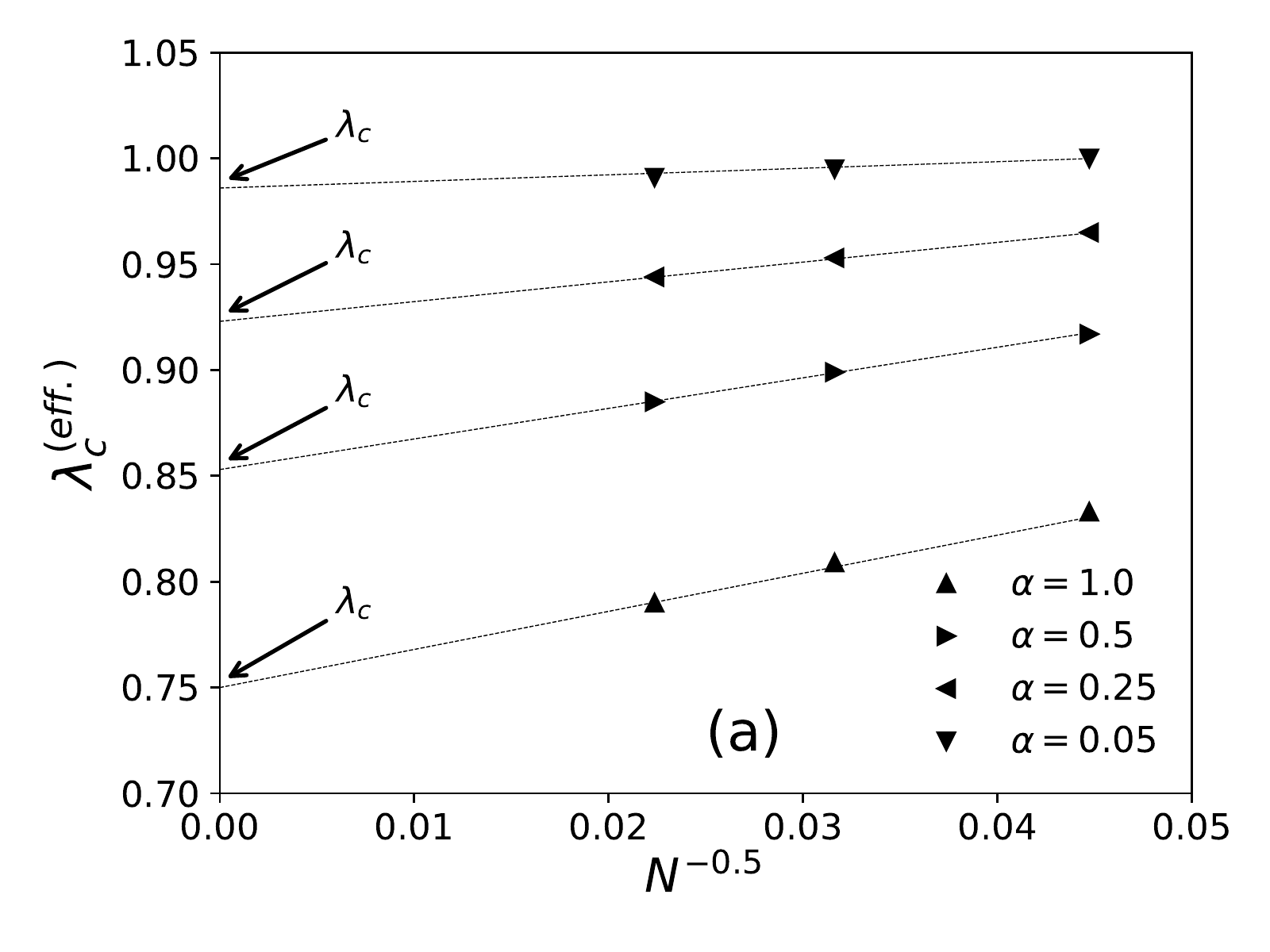} 
\includegraphics[width=6.5cm]{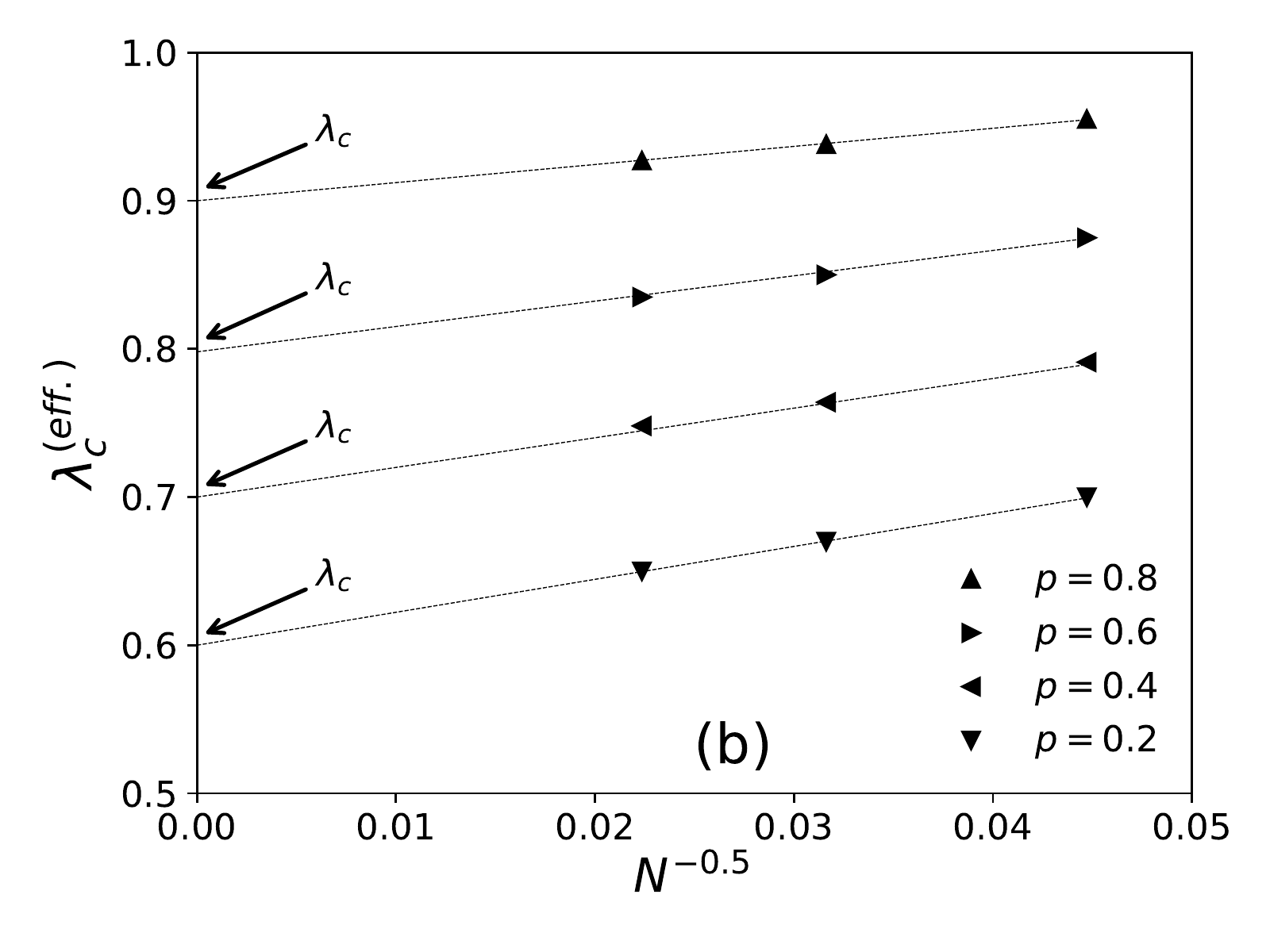} 
\end{center}
\caption{ Extrapolation study of the effective finite size critical density of agents $\lambda_c(N)$. The system size dependence is numerically fitted to  $\frac{1}{\sqrt{N}}$ and we estimate $\lambda_c$ from $\lambda_c \equiv \lambda_c(N\to \infty)$. The extrapolated values of $\lambda_c$ are $0.99,0.92,0.85,0.75$ for $\alpha = 0.05,0.25,0.5,1.0$ (strategy I), and are $0.9,0.8,0.7,0.6$ for $p = 0.8,0.6,0.4,0.2$ (strategy II).  }
\label{fig_secIV_V}
\end{figure*}

Here we consider the case where $\lambda N$
agents decide to choose among $N$ available
resources (for $\lambda$ < 1). Every day each
restaurant prepares one dish for lunch and serve
it to the visitor. If some day, any restaurant is
visited by more than one agent, then one of them
is randomly chosen and served the prepared dish;
rest leave and have to starve for that day. Thus,
every agent is required to make her choice  such
that the chosen restaurant will be visited alone by her
 (at most one agent arriving each restaurant)
to assure her lunch that day. As $\lambda$ is less
than unity here, a fraction $(1 - \lambda$) of 
restaurants will any way go vacant any day.
Additionally, a fraction $(1 - f(t))$ of restaurants
will go vacant on day $(t)$ because of overcrowding
at some restaurants due to fluctuations in choices
of the prospective customers. On any day $t$, the
average social success factor $f$ for the agents,
can be measured as

\begin{equation}\label{eq1}
f(t) = \sum_{i=1}^{N} [\delta(n_i(t))/\lambda N]
\end{equation}

\noindent
with $\delta(n) = 1$ for $n = 1$ and $\delta(n) = 0$ otherwise; $n_i(t)$ denotes the number of agents arriving at the $i$th (rank) restaurant on day $t$. $[1-f(t)]$ gives the fraction of wastage due to fluctuation of choices and $[(1 - \lambda) + (1 - f(t)]$ gives the fraction of restaurants not visited by any agent on day $t$. The goal is to achieve $f(t)$ = 1 preferably in finite convergence time ($\tau$), i.e., for $t$ $\geq \tau$, or at least as $t\to\infty$.

As usual, a dictated solution is extremely simple and efficient: A dictator asks everyone to form a queue for visiting the restaurants in order of their  respective positions in the queue and then asks them to shift their positions by one step (rank) in the next day (assuming periodic boundary condition). Everyone gets the food and the steady state ($t$-independent) social utilization fraction $f=1$. This is true even when the restaurants have ranks (agreed by all the agents or customers).

However, in democratic set-up, this dictated
solution is not acceptable and the agents or players
are expected to evolve their strategy to make the
best minority choice independently (without
presence of any dictator), using the publicly
available information about the past record of crowd
sizes in different restaurants,  such that each
arrives alone there in the respective restaurant
and gets the dish. The more successful such
collective learning, the more is the aggregated
utilization fraction $f$. Earlier studies (see
e.g.,~\cite{chakrabarti2009kolkata,ghosh2010statistics,ghosh2010kolkata,chakrabarti2017econophysics}) had proposed several learning
strategies for KPR game. Recently authors in~\cite{sinha2020phase} 
have proposed two such stochastic strategies
(strategy I and strategy II) where the  agents
collectively learn to make their decisions
utilizing the publicly available history of
crowd size  of the last day’s chosen restaurant.
Below we briefly discuss them.

\paragraph{\textbf{Strategy I:}}
On day $t$, an agent goes back to her last day's visited restaurant $k$ with probability 
\begin{subequations}
\begin{gather}
  p^{(I)}_k(t) = {[n_k(t-1)] }^{-\alpha},\:\alpha > 0. \label{eq_secIV_IA}
\shortintertext{If $n_k(t-1) > 1$, each of the $n_k(t-1)$ agents or players try to arrive at the same $k$-th restaurant next day $t$ with the above probability, and chooses a different one $(k^{'}\neq k)$ among any of the neighboring restaurants $n_r$ on day $t$, with probability }
p^{(I)}_{k^{'}}(t) = (1-p^{(I)}_k(t))/n_r. \label{eq_secIV_IB}
\end{gather}
\end{subequations}

\paragraph{\textbf{Strategy II:}}
On day $t$, an agent tries to go back to the same restaurant as chosen the earlier day $(t-1)$ with probability
\begin{subequations}
\begin{gather}
p^{(II)}_k(t) = 1,\:\textrm{if}\;n_k(t-1) = 1 \label{eq_secIV_IIA}
\;\:\textrm{and}\\
p^{(II)}_{k^{'}}(t) = p < 1,\:\textrm{if}\;n_k(t-1) > 1 \label{eq_secIV_IIB}
\end{gather}
\end{subequations}
for choosing any of the $n_r$ neighboring restaurants ($k^{'} \neq k$).

\subsection{Numerical Results}

   We  have numerically studied the steady state
dynamics of the KPR game where every day $\lambda N$
agents decide which restaurant to choose and visit
among $N$ restaurants following both the strategies I
and II. 
We consider here infinite dimensional arrangement for restaurants where the number of nearest neighboring restaurants $n_r$ to each is $(N-1)$  
and the cost to visit any of them is the same for all the time.
The maximum social utilization
$f$ obtained from Eq.(~$\ref{eq1}$) (from the point of view of
agents or players), will be denoted further by
$f^a$. Each day (iteration), parallel choice
decisions by each are  processed (following  either
strategy I or II) and used to compute $f^a$. Steady
state is identified as the state when $f^a$ does not
change (within a predefined error margin) for the
next (say, hundred) iteration.

   On day $t$, $n_i (t -  1)$ agents decide to
revisit last day’s visited restaurant (i) with
probability $p^{(I)}_k(t)$ (Eq.(~$\ref{eq_secIV_IA}$)) or
probability $p^{(II)}_k(t)$ (Eq.(~$\ref{eq_secIV_IIA}$)), else choose
any other $(k' \ne k)$ from among any of the
$(N - 1)$ neighboring restaurants for both the
strategies (Eqs.(~$\ref{eq_secIV_IB},\ref{eq_secIV_IIB}$)). After the system
stabilizes ($f^a(t)$ becomes practically independent
of $t$, the average statistics  of $f^a(t)$  are
noted  as $[f^{a(I)}]$  or $[f^{a(II)}]$  respectively
for strategies I and II. We find the power law fits for the steady state wastage fraction $(1-f^{a(I)})\sim(1-f^{a(II)})\sim(\lambda - \lambda_c(N))^{\beta}$ with $\beta = 1.0 \pm 0.05$ (see Figs.~\ref{fig_secIV_I},\ref{fig_secIV_III}) and $\tau^{(I)}\sim\tau^{(II)}\sim(\lambda_c(N) - \lambda)^{-\gamma}$ with $\gamma =  0.5 \pm 0.07$ (see Figs.~\ref{fig_secIV_II},\ref{fig_secIV_IV}) in both of the strategies I and II. Varying $\lambda$, the steady state results of $f^a$, $\tau$ for different system sizes ($N=500,1000,2000$), with $\alpha = 0.05,0.25,0.5,1.0$ in strategy I or $p=0.2,0.4,0.6,0.8$ in strategy II are considered here. All simulation are done taking maximum $N=2000$ with numbers of iteration/run of order $10^6$. For finite system sizes, the effective critical points $\lambda_c(N)$ (where $f^a$ becomes unity or $\tau$ reaches it's peak value) obtained numerically for different system sizes $(N)$ and are analyzed using finite size scaling method in Fig.~\ref{fig_secIV_V}.

It may be mentioned, that in general, for the
estimation of errors in the exponents $\beta$
and $\gamma$ in Figs.~\ref{fig_secIV_I},\ref{fig_secIV_III},\ref{fig_secIV_II},\ref{fig_secIV_IV} we tried linear fits
(without any intercept) for $\log y$ vs.
$\log x$, using best fits mostly for the
intermediate range data points for all $N$
values until they start deviating (due to
extreme fluctuations near $\lambda = \lambda_c$
and towards their saturation values for
$\lambda$ approaching unity~\cite{ghosh2010statistics,sinha2020phase})
 and anticipating their universal mean
field values in this infinite dimensional system.
From the slopes of these best fit lines for
different $\alpha$ or $p$ values, we extract
the universal exponent values and their standard
deviations. We give this higher error in the
estimate of the unified (and universal) estimate
of $\gamma$.

\subsection{Summary}

KPR is a multi-agent multi-choice repeated game where players try to learn from their past successes or failures, utilizing publicly available information on the crowd sizes at different restaurant in the past to decide which restaurant to visit that day such that she would be alone there for being served the only prepared dish. Here, asymmetric case such that $\lambda N$ $(\lambda < 1)$ agents are considered against $N$ restaurants, for sufficiently large $N$. End of each day (iteration), we have measured social utilization for agents $f^a(t) = \sum_{i=1}^{N} [\delta(n_i(t))/\lambda N]$ where $n_i(t)$ denotes number of customers visiting $i$th restaurant on day $t$.

As shown in Fig.~\ref{fig_secIV_I} (for strategy I) and
Fig.~\ref{fig_secIV_II} (for strategy II), the social wastage
fraction $(1 - f^a)$  vanishes at the effective
critical point $\lambda_c(N)$ with the critical
exponent $\beta$ value near unity. Also, from
Fig.~\ref{fig_secIV_III} (for strategy I) and Fig.~\ref{fig_secIV_IV} (for strategy
II), we see that the the convergence or relaxation
time $t$, required for $f^a$ to stabilize,
divergence near the same critical points
$\lambda_c(N)$ for the respective strategies,
with the exponent $\gamma$ value about 1/2.
Additionally, the finite size scaling analysis
$\lambda _c(N) \sim  \lambda_c + const. N^{-1/(d\nu)}$,
where $\lambda_c$ corresponds to $\lambda_c(N)$ for
$N$ going to infinity and $d$ corresponds to the
effective dimension, suggests the effective correlation
length exponent $d\nu$ value to be around 2 for both
the strategies, as expected for such mean field
(infinite range systems).

In~\cite{sinha2020phase}, we have studied the dynamics of the KPR
game following the same two strategies for the
case $\lambda = 1$. For $\lambda = 1.0$, where the
critical points $\lambda_c $ (for both the
strategies) vanish, the universality class (values
of the critical exponents $\beta$  and $\gamma$
were observed to be distinctly different, and this
point needs further investigations. We may, however,
note that since at $\lambda = 1$, the number of
both agents and restaurants are same ($N$), full
social utilization (where $f^a = 1 = f^r$, occurring
at $\alpha = 0_+$ for strategy I and at $p = 1_-$
for strategy II) induce an additional frustrating
constraint in the collective choice dynamics
involved here.

\begin{figure}[H]
\begin{center}
\includegraphics[width=15cm]{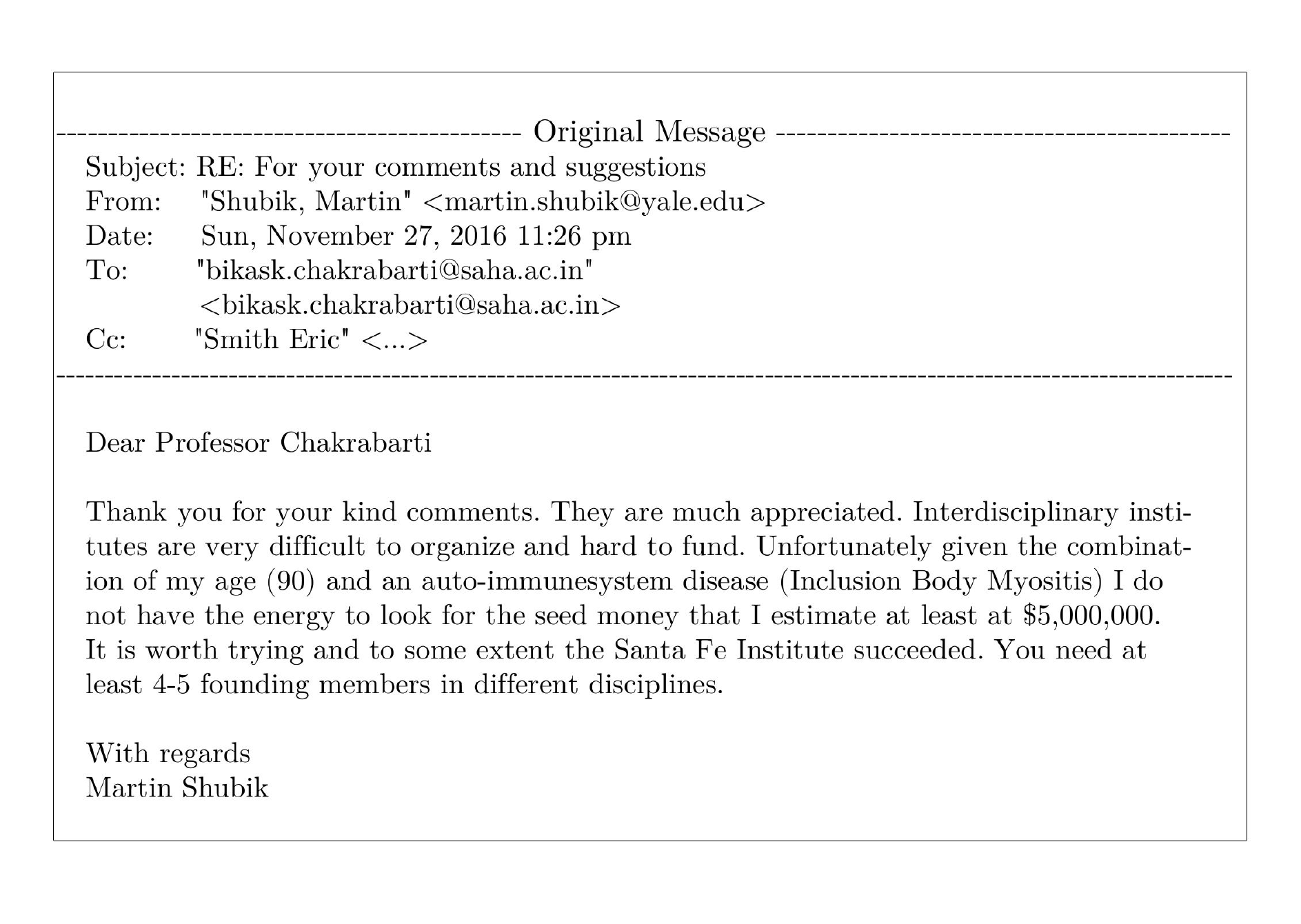}
\end{center}
\caption{ The first part of the email
conversation  between (Late) Martin Shubik and
BKC. Second part (email from BKC; appended to
this part) is continued in Fig.~\ref{fig_secV_partII}. The precise
suggestions made in this immediate response
indicate Shubik's prior plan for such
`interdisciplinary institutes' in economics.
}
\label{fig_secV_partI}
\end{figure}

The KPR game models have been extended already and
used to study real life problems like resource allocation
in Internet of Things~\cite{park2017kolkata}, vehicle for hire~\cite{martin2019}, or matching
in mobility markets~\cite{martin2017vehicle}, and etc. We hope, the KPR game
models will be utilized much more effectively in the context
of much wider practical areas of collective learning dynamics
and choices.

\section{Future of Econophysics: Some Perspective}

One often says that the main purpose of
economic activity is to optimize the
limited funds of labor and capital, natural
and technical resources and capital resources,
to satisfy our (practically) unlimited needs.
``Economic science is therefore the science of
efficiency, and as such, it is a quantitative
science.''~\cite{allais1968economics} (see also~\cite{frey1999economics}). We have already
argued~\cite{chakrabarti2016can} in section 2 that epistemologically
economics belongs to natural  science (and
not mathematics). It begins with observation
which are to be analyzed using logic or
mathematics and eventually should end in
observation, as in all natural sciences.
Since 1990s, most Universities of the world
offer Science Graduation degrees (Bachelor of
Science or Master of Science degrees) in
economics (in addition to Bachelor of Arts
or Master of Arts from Fine Arts or Humanities
Departments).

Robert Solow~\cite{solow1997did} pointed out that in the 1940s, economics had been
basically a descriptive and institutional subject for a `gentleman
scholar'. The textbooks of those days were `civilized' and discursive. ...
``Formal analysis were minimal and it made economics the domain of intuitive
economists''. He concluded his summary of the state of economics near the
end of the 20th century ``with a paraphrase of Oscar Wilde’s description of
a fox hunt - `the unspeakable in pursuit of the inedible' - saying that
perhaps economics was an example of `the over-educated in pursuit of the
unknowable.'.''~\cite{solow1997did}. Despite the ongoing controversies today in the
field of economics the ``New Millennium economists are far more comfortable
with what they do after the changes in the structure  and content of
economics over the last half century.''~\cite{colander2000new}. The root cause of these
changes have been identified by Colander~\cite{colander2000new} to be due to the rise of
Complexity Science since early 1980s. In fact, concepts from physics had
continually been absorbed into the main stream economic formulation of
ideas and models. As Venkat Venkatasubramanian noted in his recent 
book~\cite{venkatasubramanian2017much}, ``Concepts such as equilibrium, forces of supply and demand,
and elasticity reveal influence of classical mechanics on economics. The
analytical model of utility-based preferences can be traced back to Daniel
Bernoulli, the great Swiss mathematical physicist from nineteenth century.
One of the founders of neoclassical economics, Irving Fisher, was trained
under the legendary Yale physicist, Jisiah Willard Gibbs, a co-founder of
the discipline of statistical mechanics. Similarly, Jan Tinbergen, who
shared the first Nobel Prize in Economics in 1969, was the doctoral
student of the great physicist Paul Ehrenfest at Leiden University.''

Indeed, more specifically as discussed in section
2, we would like to correlate these changes to occur
following the successful development in econophysics of
the Simulated Annealing technique~\cite{kirkpatrick1983optimization} in 1983 for
Traveling Salesman type multi-variable optimization
problems, and  other successive developments in
econophysics of analyzing correlations in  stock
prices (see e.g.,~\cite{mantegna1991levy,stanley2000introduction}) or the kinetic
exchange modelings of income and wealth
distributions (see e.g.,~\cite{yakovenko2009colloquium,chakrabarti2013econophysics}). The
statistical physics of TSP, as an example of successful developments in econophysics, had already 
been introduced in our 2010 econophysics
textbook~\cite{sinha2010econophysics}, which has been the only `suggested textbook'
for a decade (since inception in 2012) for the formal
course on econophysics, offered (by Diego Garlaschelli) at
the Physics Department of the Leiden University
(see the course prospectus for 2012-13 through that
of 2021-22~\cite{leidenuniv}), where one of the first Nobel-laureates in economics
Jan Tinbergen came from.

Econophysics has come as an exceptional development in
interdisciplinary sciences (see e.g.,~\cite{dash2019story} for a popular
exposition on this development). Historically economics, more
specifically social sciences, belonged to the Humanities departments
and not of Science. For earlier interdisciplinary developments of
Astrophysics, Biophysics or Geophysics, the  scenario and ambiance had
been quite different. The mother departments had been parts of the same
science schools and even the corresponding resources like books, journals, 
and also the faculty had strong overlaps and could be shared. The
marriage negotiations for Econophysics have been difficult, though
extremely  desirable and natural; as the saying goes: ``marriage between the
King of natural sciences with the Queen of social sciences!''

Regular interactions and collaborations between the communities of
natural scientists and social scientists, are however rare even today! Though, as
mentioned already, interdisciplinary research papers on econophysics and
sociophysics are regularly being published at a steady and healthy rate,
and a number of universities (including Universities of Bern, Leiden,
London, Paris and Tufts University) are offering the interdisciplinary
courses on
econophysics and sociophysics, not many clearly designated professor
positions, or other faculty positions for that matter, are available yet
(except for econophysics in Universities of Leiden and London). Neither
there are designated institutions on these interdisciplinary fields, nor
separate departments or centres of studies for instance. Of course there
have been several positive and inspiring attempts and approaches from both
economics and finance side (see e.g.,~\cite{shubik2016guidance,jovanovic2017econophysics},
along with a number of those~\cite{pareschi2013interacting,ribeiro2020income,richmond2013econophysics,slanina2013essentials,aoyama2017macro} from physics which have already been appreciated in
the literature. Indeed the thesis~\cite{schinckus2018physics} in August 2018, Department of
History and Philosophy of Science, University of Cambridge,  by financial
economist Christophe Schinckus (one of the co-editors of this special
issue), says that ``In order to reconstruct the subfield of econophysics, I
started with the group of the most influential authors in econophysics and
tracked their papers in the literature using the Web of Science database
of Thomson-Reuters (The sample is composed of: Eugene Stanley, Rosario
Mantegna, Joseph McCauley, Jean-Pierre Bouchaud, Mauro Gallegati, Benoît
Mandelbrot, Didier Sornette, Thomas Lux, Bikas Chakrabarti and Doyne
Farmer). These key authors are often presented as the fathers of
econophysics simply because they contributed significantly to its early
definition and development. Because of their influential and seminal
works, these scholars are actually the most quoted authors in
econophysics. Having the 10 highest quoted fathers of econophysics as a
sample sounds an acceptable approach to define
bibliometrically the core of econophysics.” Also,
the entry on `Social Ontology' in The Stanford
Encyclopedia of Philosophy~\cite{epstein2018social}, as discussed in
section 2, confirms positive impact of such
econophysics and sociophysics researches on the
overall modern philosophical outlook of social
sciences.

\begin{figure}[H]
\begin{center}
\includegraphics[width=14.2cm]{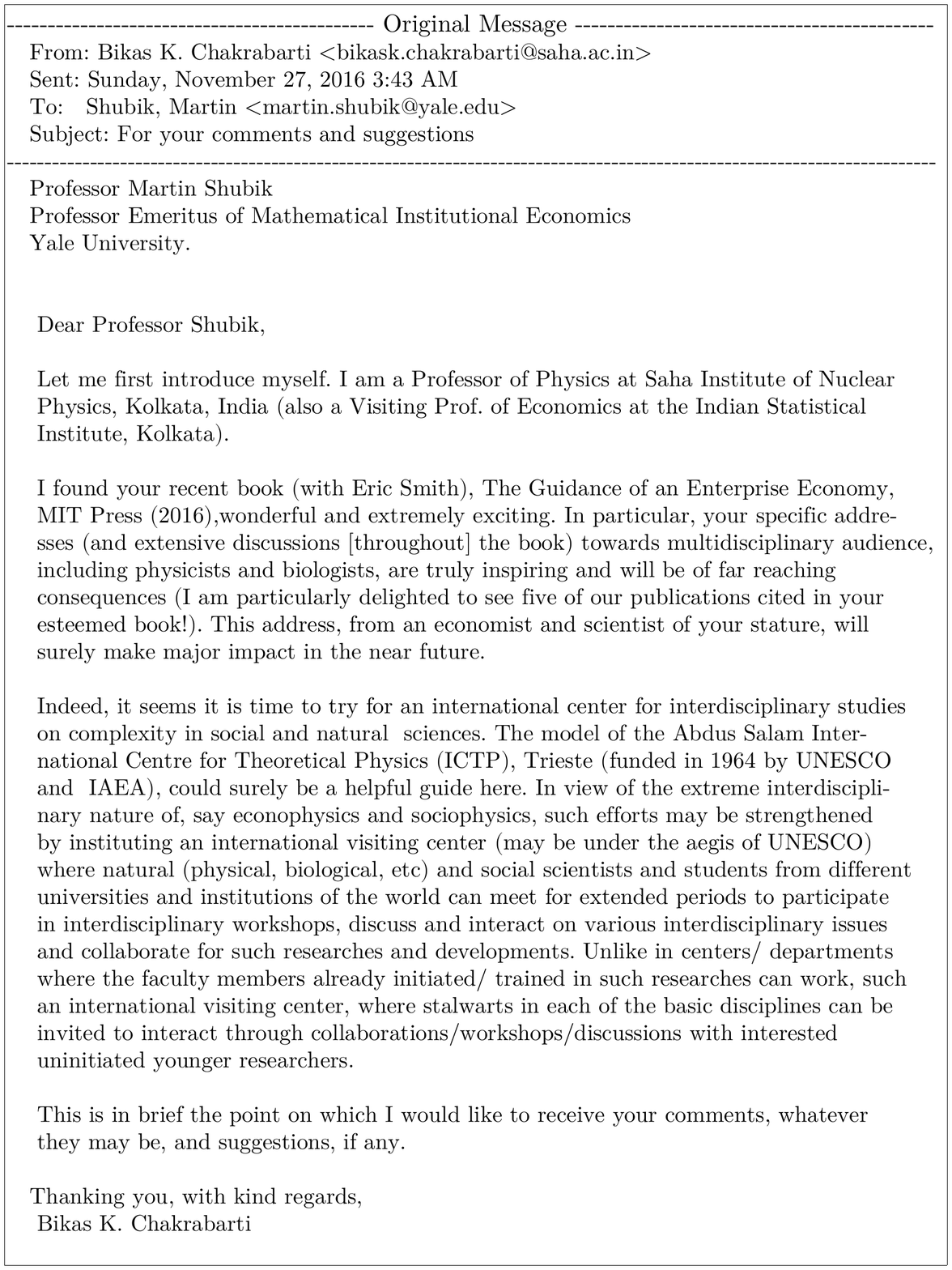}
\end{center}
\caption{Email conversation in the end of 2016
between (Late) Martin Shubik and BKC regarding
interdisciplinary developments in economics and
the possibility of  setting up an International
Center for Interdisciplinary Studies on Complexity
in Social Sciences. This email from BKC was
appended to the response email (Fig.~\ref{fig_secV_partI}) from
Shubik. The (Yale) date and time mark in the
 mail-header (and that for BKC's in Fig.~\ref{fig_secV_partI}, on arrival in Kolkata) indicate hardly any time
gap between the  two and the readiness with the
precise suggestions indicate Shubik’s prior
thinking in similar line.
}
\label{fig_secV_partII}
\end{figure}

We may note however, a recently published highly
acclaimed massive (580 page) book~\cite{shubik2016guidance} on economics
(`landmark volume', said E. Roy Weintraub, `creative,
elegant and brilliant work', said  W. Brian Arthur
and `written by master economists', said D. Colander)
by (Late) Martin Shubik (Ex-Seymour Knox Professor
Emeritus of Mathematical Institutional Economics,
Yale University and Santa Fe Institute) and Eric
Smith (Santa Fe Institute) discussed extensively
on econophysics approaches and in general on the
potential of interdisciplinary researches inspired
by the developments in natural sciences. Getting
somewhat excited, I wrote to Martin Shubik in late
2016 that their  book can also serve as an
outstanding `white-paper' document in favor
of a possible Proposal for an International
Center for Interdisciplinary Studies on Complexity
in Social Sciences. He immediately responded and
gave his impression about the difficulties involved
and indicated very briefly about the minimal
financial and structural requirements (both my
letter to him and his response is appended below
(Figs.~\ref{fig_secV_partI},\ref{fig_secV_partII}).

This ready and specific comments by Shubik clearly
suggests that he actually had thought about the
 need of such an International Center 
for fostering interdisciplinary research which need
to be more inclusive than for  example the Santa
Fe Institute. The model of the Abdus Salam
International Centre for Theoretical Physics (ICTP),
Trieste (funded by UNESCO and IAEA), was considered to provide helpful guidance for us here. It was 
contemplated, if an ICTP-type interdisciplinary
research institute could be initiated for researches
on econophysics and sociophysics (see e.g,~\cite{chakrabarti2019international}).
Though Shubik (who died in 2018 at the age of 92)
agreed also to be one of its founding members, we
could not make any progress yet. We may also note
that Dirk Helbing and colleagues have been trying
for an European Union funded `Complex Techno-Socio-
Economic Analysis Center' or `Economic and Social
Observatory' for the last decade or so  (see
Ref.~\cite{helbing2011understanding} containing the White Papers arguing for
their proposed project). We are also aware that Indian
Statistical Institute had taken a decision to initiate
a similar Centre in India (see `Concluding Remarks'
in~\cite{ghosh2013econophysics}).

Hope, some such international visiting centers will come up
soon and with them the spread of such interdisciplinary ideas will achieve more coherence and will lead to major success in such
researches.

\section*{Acknowledgement}
We acknowledge all our colleagues (mentioned by name in sec. 2) for the collaborations. BKC is grateful to J.C. Bose National Fellowship (DST, Govt. of
India) grant for support.

%%%%%%%%%%%%%%%%%%%%%%%%%%%%%%%%%%%%%%%%%%
\end{document}